\documentclass[acmsmall,review=false,screen,table,nonacm]{acmart}

\AtBeginDocument{%
  \providecommand\BibTeX{{%
    \normalfont B\kern-0.5em{\scshape i\kern-0.25em b}\kern-0.8em\TeX}}}

\usepackage{placeins}
\usepackage{xcolor}
\usepackage{footnote}
\usepackage{enumitem} 
\usepackage{enumerate}
\usepackage{multirow}
\usepackage{afterpage}
\usepackage{longtable}
\usepackage[compact]{titlesec}

\begin{document}
\raggedbottom


\newlist{inlineroman}{enumerate*}{1}
\setlist[inlineroman]{itemjoin*={{; and }},afterlabel=~,label=(\roman*)}

\newcommand{\inlinerom}[1]{
\begin{inlineroman}
#1
\end{inlineroman}
}
\newenvironment{myindentpar}[1]%
 {\begin{list}{}%
         {\setlength{\leftmargin}{#1}}%
         \item[]%
 }
 {\end{list}}
\newboolean{showcomments}
\setboolean{showcomments}{true} 
\ifthenelse{\boolean{showcomments}}
  {\newcommand{\nb}[2]{
    \fcolorbox{black}{yellow}{\bfseries\sffamily\scriptsize#1}
    {$\blacktriangleright$#2$\blacktriangleleft$}
   }
   \newcommand{\version}{\emph{\scriptsize$-$working$-$}}
  }
  {\newcommand{\nb}[2]{}
   \newcommand{\version}{}
  }
  
\newcommand\Ines[1]{\nb{Ines}{\textcolor{blue}{#1}}}
\newcommand\sabrina[1]{\nb{Sabrina}{\textcolor{red}{#1}}}

\title{Usage Control Specification, Enforcement, and Robustness: A Survey}\thanks{This work is funded by the European Union Horizon 2020 research and innovation programme under the Marie Skłodowska-Curie grant agreement No 860801. Sabrina Kirrane is funded by the FWF Austrian Science Fund and the Internet Foundation Austria under the FWF Elise Richter and netidee SCIENCE programmes as project number V 759-N}


\author{Ines Akaichi} \orcid{0000-0002-6020-5572}
\email{ines.akaichi@wu.ac.at}

\author{Sabrina Kirrane} \orcid{0000-0002-6955-7718}
\email{sabrina.kirrane@wu.ac.at}
\affiliation{%
  \institution{Institute for Information Systems \& New Media, Vienna University of Economics and Business}
  \city{Vienna}
  \country{Austria}
}

\renewcommand{\shortauthors}{Akaichi and Kirrane}

\begin{abstract}
The management of data and digital assets poses various challenges, including the need to adhere to legal requirements with respect to personal data protection and copyright. Usage control technologies could be used by software platform providers to manage data and digital assets responsibly and to provide more control to data and digital asset owners. In order to better understand the potential of various usage control proposals, we collate and categorize usage control requirements, compare the predominant usage control frameworks based on said requirements, and identify existing challenges and opportunities that could be used to guide future research directions.
\end{abstract}


\maketitle

\section{Introduction}
Modern decentralized systems, such as the \emph{Internet of Things (IoT)} and \emph{virtual data spaces}, face a variety of challenges from a data and digital asset management perspective. According to \citet{Zrenner2019}, data owners are reluctant to share their data with decentralized systems, as often they have no control over how their data are used. 
Since the \emph{General Data Protection Regulation} (GDPR) \cite{EUdataregulations2018} entered into force, in 2018, the need to provide more control and transparency to data subjects with respect to how personal data are \emph{collected}, \emph{stored}, and \emph{processed} is mandated in Europe, and also outside of the European Union if the data relates to European citizens. More broadly, the importance of digital asset management is underlined by the \emph{new copyright legislation} \cite{EUcopyrightregulations2021}, which came into effect in 2021, with the aim to protect creativity in the digital age. 

When it comes to digital asset management, \citet{Pretschner2009ANOO} and \citet{10.1145/984334.984339} highlight that the sharing of data in decentralized environments goes beyond traditional access control, as existing solutions do not provide control over data usage once access to the data has been granted. 
Technologies that aim to address this challenge, which are usually classified as usage control \cite{pretschner2006distributed}, aim to ensure that data consumers handle data according to usage rules stipulated by data owners. 
More broadly, usage control is an umbrella term for data management software that caters for data protection, copyright, and/or various legislative and institutional policies \cite{Pretschner2009ANOO}. 

The term \emph{usage control} was first introduced by \citet{10.1145/984334.984339} whose research focuses on supporting the continuous monitoring of digital asset usage in dynamic distributed environments. Over the years, researchers have proposed various usage control conceptual models (cf., \cite{CAO2020998,10.1145/984334.984339,Neisse2015SecKitAM}) and policy languages (cf., \cite{10.1007/978-3-540-74835-9_35,SPL2018,10.1007/978-1-4419-6794-7_11}). Other works focus on enforcing the respective policies, via proactive mechanisms that aim to prevent policy violations (cf., \cite{Jung2014,Lazouski2012APF,DBLP:conf/codaspy/KumariPPK11}) and reactive mechanisms that detect security breaches and policy violations (cf., \cite{basin2011monpoly,10.1007/978-3-540-74409-2_11,Etalle2007}).
Additionally, there are a handful of surveys that aim to better understand the state of the art with respect to usage control. For instance, \citet{DBLP:journals/ieeesp/PretschnerHSSW08} survey existing usage control mechanisms with a specific focus on \emph{digital rights management} (DRM) technologies. 
Another survey conducted by \citet{Lazouski2010UsageCI} focuses specifically on the usage control (UCON) model proposed by \citet{10.1145/984334.984339}, with the authors reviewing the various implementations and extensions. In turn, \citet{DBLP:conf/IEEEares/Nyre11} analyzes the strengths and weaknesses of existing usage control solutions, with a particular focus on enforcement mechanisms.

Considering the potential of usage control as a tool for ensuring compliance with respect to data protection, copyright, as well as institutional policies, there is need for a more holistic overview of existing works and their support for a broad set of usage control requirements. 
Towards this end, in this paper, we perform a comprehensive analysis of existing usage control frameworks based on a variety of usage control requirements gleaned from the literature. Our primary contributions are summarized as follows: \inlinerom{ \item we present and align various usage control concept definitions found in the literature; \item we propose a comprehensive taxonomy of usage control requirements that are commonly used to guide the development of usage control solutions; \item we conduct a qualitative comparison of the predominant usage control proposals found in the literature based on the aforementioned taxonomy \item we draw on the comparison in order to outline various challenges and opportunities for the usage control domain in general and decentralized systems in particular.} 

The remainder of this paper is structured as follows:
In Section 2, we present our motivating use case scenario and introduce several important usage control concepts. In Section 3, we discuss the methodology underpinning our literature review and requirements taxonomy generation process. In Section 4, we describe the proposed taxonomy of usage control requirements. In Section 5, we compare and contrast the predominant usage control frameworks found in the literature. In Section 6, we discuss open challenges and opportunities for the usage control domain. Finally, we conclude by summarizing the paper in Section 7.

\section{Motivational Scenario}
We start by describing a practical usage control scenario that can be used to guide our analysis. Following on from this, we provide the necessary background in terms of usage control concept definitions.

\subsection{Use Case}
\begin{figure}[t!]
  \centering
  \includegraphics[scale=.5]{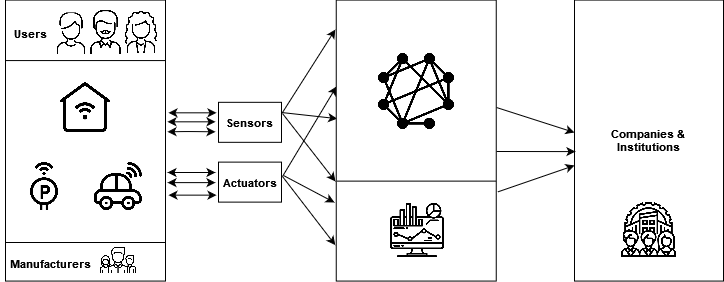}
  \caption{Descriptive Usage Control Scenario}
  \label{fig:useCase}
\end{figure}

Our usage control scenario is inspired from the \emph{internet of things} (IoT) domain.
Figure \ref{fig:useCase} illustrates a smart city scenario, where residents make use of multiple smart objects, such as  smart homes, cars, parking lots, watches, etc. Smart objects contribute, among other things, to simplifying every day activities. These objects produce different types of data that are captured by sensors or actuators. In such a scenario, data can relate to the smart objects, such as power consumption, battery status, etc., or the users themselves, such as Global Positioning System (GPS) location or any other type of private information that relates specifically to the user. 
Different stakeholders, such as institutions that manage the supply of water or energy, are interested in the data produced by these smart objects in order to derive insights on consumption that can be used to optimize their service offerings. Such information could also be used by marketing companies in order to devise new or adjust existing marketing strategies. Thus, the manufacturers of these objects may host or use data sharing platforms whereby data resulting from the use of smart objects are shared with both their customers and various third parties. Such platforms could offer the following sharing possibilities to subscribers: \inlinerom{\item an option to download data relating to smart objects or their users; \item the ability to access ad-hoc analysis and statistics about specific smart objects \item the possibility to perform on-the-fly analysis based on statistical or machine learning models.}
The decentralized nature of this data sharing scenario and the nature of the data imply usage concerns that involve personal data (e.g., control the usage of location information) and regulations (e.g., delete all personal data after a certain time, as mandated by the GDPR), among others.

\subsection{Usage Control: What and How?}
The goal of this section is to establish a common understanding in terms of the various concepts that are collectively used to specify usage control policies (what?) and to describe how usage control policies can be enforced (how?).
\subsubsection{What?}
\citet{pretschner2006distributed} refer to an entity that provides a resource (e.g., data) together with a policy that describes its access and usage restrictions as a \emph{data provider}. Whereas, an entity that receives a copy of a resource and the respective policy is called a \emph{data consumer}.
\citet{10.1145/984334.984339} refer to both the data consumer and the data provider as \emph{subjects}. 
The resources that are governed by the policy are referred to as \emph{data} by \citet {pretschner2006distributed} and \emph{objects} by \citet{10.1145/984334.984339}. 
Usage control policies express restrictions in the form of rules that govern the use of resources, in particular the \emph{actions} or \emph{operations} on the data that can be performed by data consumers (e.g., accessing, processing, downloading). \citet{10.1145/984334.984339} refer to such actions as \emph{rights}. 

Furthermore, \citet{bettini2003provisions} and \citet{pretschner2006distributed} define usage control rules in terms of two basic classes: \emph{provisions} and \emph{obligations}. Provisions refer to specific activities that need to be performed before an access decision is taken (i.e., before access to data is provided), while obligations refer to activities that need to be performed in the future (i.e., after access to data has been granted). The problem with this definition is that it does not capture the generality of usage rules in terms of what should or should not (i.e., positive and negative obligations) and what can or can not (i.e., permissions and prohibitions) be done with the data \cite{Garcia2005}. This definition focuses only on the specific activities that must be performed by users before and after access to the data is provided. 
%
\citet{10.1145/984334.984339}, in turn, define usage control policies in terms of decision factors that determine the final decision made by a system, also known as the usage decision.
Decision factors denote: \emph{authorizations}, \emph{obligations}, and \emph{conditions}.
Authorizations refer to constraints on subject or object attributes that are used to enforce usage decisions. Attributes denote the properties or the capabilities of subjects or objects. For instance, based on our use case scenario, we could specify an attribute called \emph{role} that could take the following values: \textit{admin} for platform operators, \textit{owner} for smart device owners, and \textit{external} for third parties. Additionally, we could define a \emph{subscription} attribute that indicates whether a stakeholder has a subscription to a data sharing platform or not.
Another important aspect with respect to attributes is the level of sensitivity attached to the data, which may be different depending on the type of data, e.g., energy consumption, location, medical, etc. An example of such a rule could be that \emph{full access to personal data that is classified as sensitive should only be given to the data subject themselves}. 
While, obligations refer to activities that an entity must carry out in order to be permitted to perform particular actions.
For example, in our motivating scenario, we assume that a stakeholder, namely, the marketing company, is interested in performing data analysis with respect to electricity consumption, however \emph{if the marketing company wishes to download user data (with the users consent) the company must delete the data within 10 days}. Conditions, in turn, refer to environmental or system requirements that have to be satisfied in order to perform certain actions on data objects. For example, a condition can refer to the purpose for which the data may be used, for instance \emph{the marketing company may only use data for scientific purposes}. Conditions are also used to refer to contextual information, such as the \emph{time} and \emph{location} of environmental conditions.

Moreover, \citet{10.1145/984334.984339} present an important aspect of usage control, which is \emph{the continuity of enforcement}.
This feature implies that usage decisions are enforced not only when data providers or data consumers generate access requests, but also during the ongoing usage of the data. Continuity of enforcement also implies that a usage control system must continuously evaluate conditions and obligations. This means, that the conditions have to be satisfied before (i.e., pre-condition) or during (i.e., ongoing-condition) a usage process. In addition, the  usage control system must ensure the fulfillment of obligations before (i.e., pre-obligation), during (i.e., ongoing-obligation), and after (i.e., post-obligation) data are accessed.
Once the conditions no longer hold or the obligations are not met by data consumers, the system can deny or revoke access to the data.
In this paper, we define usage control policies in terms of deontic concepts, both due to their generality \cite{Ortalo1996UsingDL} but also because they provide support for what should or should not as well as what can or cannot be done with data  \cite{Ortalo1996UsingDL,bettini2003provisions}. This definition is close to the definition of \citet{10.1145/984334.984339}, but includes more specific decision factors (i.e., \emph{permissions}, \emph{prohibitions}, \emph{obligations}, and \emph{dispensations}) that are needed in order to represent legislative requirements. 
Bearing these different definitions in mind, usage control policies can be defined in terms of the continuity of enforcement and the following decisions factors: \emph{permissions}, \emph{prohibitions}, \emph{obligations}, \emph{dispensations}, \emph{conditions}, each of which support a variety of \emph{attributes}.
Given that \emph{obligations}, \emph{conditions}, and \emph{attributes} have already been defined by \citet{10.1145/984334.984339}, we build upon the definitions and examples presented above by providing definitions and examples for the remaining concepts.
\emph{Permissions} represent positive authorizations that allow entities to perform actions. For instance, \emph{the marketing company is permitted to download data about the energy consumption of a specific neighborhood}. Prohibitions refer to negative authorizations, implying that an entity is not allowed to perform the specified actions. For example, \emph{the marketing company is prohibited from downloading personal information}. Dispensations refer to  actions that an entity is no longer required to perform, thus, they act as waivers for existing obligations. For example, \emph{a user is exempt from deleting their data after usage because they are the owner}. 

\begin{figure}[t!]
  \centering
  \includegraphics[scale=.5]{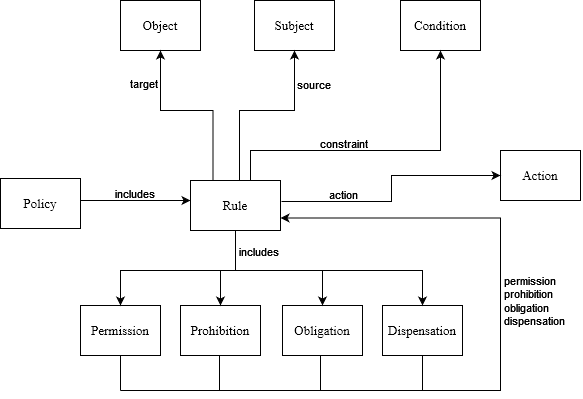}
  \caption{A Usage Control Model based on Deontic Concepts}
  \label{fig:UsageControModel}
\end{figure}

\begin{figure}[t!]
  \centering
  \includegraphics[width=390pt]{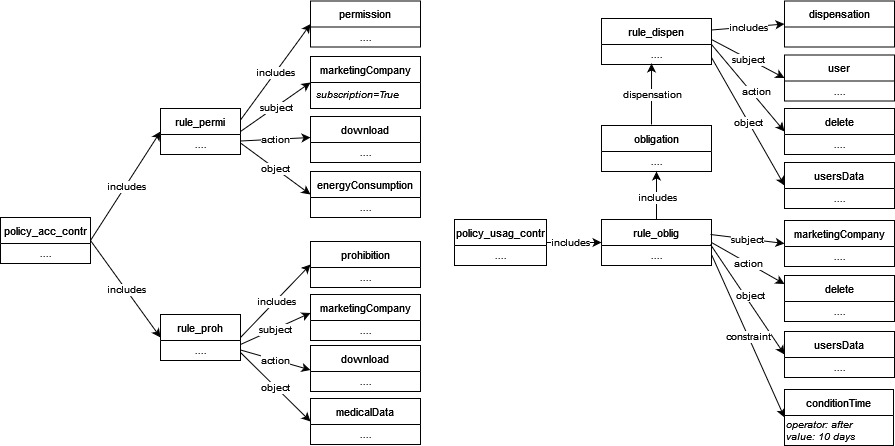}
  \caption{Usage Control Policies based on our Motivating Use case Scenario}
  \label{fig:UsageControModelInstances}
\end{figure}

Figure \ref{fig:UsageControModel} depicts a usage control model that encapsulates the various decision factors that are necessary in order to create a usage control policy based on the given definition.
In the proposed model, a policy is made up of a set of rules that encode \texttt{permissions}, \texttt{prohibitions}, \texttt{obligations}, or \texttt{dispensations}. Each rule is associated with an \texttt{action} that is performed by a \texttt{subject} on a target \texttt{object}. A rule can also be constrained by one or more \texttt{conditions}.
The various entities in the model can have specific attributes (e.g. policy attributes and condition attributes). 
In addition, the model supports \emph{nested rules} that can express nested requirements, which are needed to encode regulatory requirements, such as those set forth by the GDPR.
The modeling of the nested rules is inspired by the \emph{open digital rights language} (ODRL) regulatory profile proposed by \citet {DeVos2019ODRLPM}.
Figure \ref{fig:UsageControModelInstances} shows an instantiation of the model using a permission, a prohibition, and an obligation with a nested dispensation that are inspired by our motivating use case scenario.

\subsubsection{How?}
According to \citet{DBLP:journals/ieeesp/PretschnerHSSW08}, usage control policies immediately raise the question of enforcement. Indeed, policies depend on the actual implementation and deployment of a usage control solution, which is limited to the ability of the solution to continuously validate and enforce  the usage control decisions \cite{pretschner2006distributed,10.1145/984334.984339}.
In this paper, we refer to a usage control framework, as a complete framework that allows for the specification, the enforcement, and the administration of usage policies. According to \citet {Zhang2008TowardAU}, a usage control framework addresses both the ``how'' and ``what'' aspects of policy enforcement. Generally speaking, usage control frameworks are comprised of the following components: 
\inlinerom{\item a formal machine-readable policy language that is used to express usage control policies; \item an enforcement mechanism that can monitor compliance with said policies \item an administration interface that can be used to manage and monitor usage control policies.} 
Although there are several frameworks in the literature, a detailed analysis is needed in order to better understand how well the various usage control proposals support various usage case requirements, which is the aim of this paper.

\section{Methodology}
Our survey was guided by the integrative literature review methodology proposed by \citet{Torraco2005}. While, the requirements taxonomy generation process followed the development methodology suggested by~\citet{doi:10.1057/ejis.2012.26}. Finally, the requirements taxonomy was used to perform a targeted analysis of the predominant usage control frameworks found in the literature.

\subsection{Literature Identification}

The literature review involved the identification of concrete research questions (\emph{RQs}) and the corresponding review strategy \citep{Torraco2005}. Considering that our overarching goal was to assess the status quo in terms of usage control solutions found in the literature, and to identify open challenges and opportunities, our research was guided by the following research questions:\\
 \emph{RQ1.} What requirements are used to guide the development of usage control solutions?\\
 \emph{RQ2.} Which specific requirements are supported by the predominant usage control frameworks found in the literature?\\
 \emph{RQ3.} What are the open challenges and opportunities for the usage control domain?

\smallskip\noindent
The papers subject to the review were found using the \textit{Google Scholar} search interface using keywords consisting of a combination of ``usage control'', ``requirements'', and ``framework'' and the following inclusion and exclusion criteria:
\begin{inlineroman}
\item include papers that mention usage control requirements;
\item include papers that propose usage control frameworks and/or extensions
\item  exclude papers that focus on non-implemented frameworks.
\end{inlineroman}

First, we collected all papers that were returned using our Google Scholar keyword search. 
Following on from this, the title, abstract, introduction, and conclusion of all articles that matched our keyword search were examined for relevancy based on our inclusion and exclusion criteria. Then, a full read through of the remaining articles was performed in order to identify and extract usage control requirements (\emph{RQ1}) and to compare and contrast existing frameworks (\emph{RQ2}) as well as derive open challenges and opportunities (\emph{RQ3}).
For every paper, an iterative backward (i.e., searching the citations of the identified articles) and forward (i.e., locating the papers that cite the identified articles) search was performed in order to improve the coverage of related work. 

\begin{table}[t]
\centering
\scriptsize
\caption{Usage Control Requirements and Sources}
\resizebox{\textwidth}{!}{%
\begin{tabular}{lccccccccccccc}
\rowcolor[HTML]{C0C0C0} 
 &
  \multicolumn{4}{c}{\textbf{Specification}} &
  \multicolumn{5}{c}{\textbf{Enforcement}} & 
  \multicolumn{4}{c}{\textbf{Robustness}} \\ 
  \rowcolor[HTML]{C0C0C0} 
   & & & & & & & & &   &  & &  & \\
   \rowcolor[HTML]{C0C0C0} 
\textbf{Authors} &
  Expressiveness &
  \begin{tabular}[l]{@{}c@{}}Flexibility \& \\ Extensibility\end{tabular} &
  \multicolumn{1}{l}{Unambiguous} &
  \multicolumn{1}{l}{\begin{tabular}[c]{@{}c@{}}Formal \\ Semantic\end{tabular}} &
  Preventive &
  Detective &
  \multicolumn{1}{l}{\begin{tabular}[c]{@{}c@{}}Continuity \\ of enforcement\end{tabular}} &
  \multicolumn{1}{l}{\begin{tabular}[c]{@{}c@{}}Conflict Detection\\  \& Resolution\end{tabular}} &
  Administration &
  \multicolumn{1}{l}{\begin{tabular}[c]{@{}c@{}}Interoperability\\  \& Compatibility\end{tabular}} &
  \multicolumn{1}{l}{\begin{tabular}[c]{@{}c@{}}Performance\\  \& Scalability\end{tabular}} &
  Usability &
  Reliability \\ 
   & & & & & & & & &  & &   &  & \\
\citet{Kattinproceedings2008} &
  X &
   &
   &
   &
   &
   &
   &
   &
   &
   &
   &
   & 
  \multicolumn{1}{l}{} 
  \\ 
   \rowcolor[HTML]{EFEFEF} 
 & & & & & & & & &  & &   &  & \\
  \rowcolor[HTML]{EFEFEF} 
\citet{DBLP:conf/icsnc/HiltyPSSW06} &
  X &
   &
   &
   &
   &
   &
   &
   &
   & 
   & 
   &X
   & 
  \\ 
   & & & & & & & & &  & &   &  & \\
\citet{Kumari2010RequirementsAF} &
  X &
   &
   &
   &
   &
   &
   &
   &
   & 
   & 
   &X
   & X
  \\ 
   \rowcolor[HTML]{EFEFEF} 
   & & & & & & & & &  & &   &  & \\
    \rowcolor[HTML]{EFEFEF} 
\citet{DBLP:journals/ieeesp/PretschnerHSSW08} &
  X &
   &
   &
   & X
   &
   &
   &
   &
   &X
   &
   &
   & X 
   \\ 
   & & & & & & & & &  & &   &  & \\
\citet{pretschner2006distributed} &
  X &
   &
   &
   &
   & X
   & X
   &
   &
   &
   &
   &
   &
   X \\ 
   \rowcolor[HTML]{EFEFEF} 
   & & & & & & & & & &  &   &  & \\
    \rowcolor[HTML]{EFEFEF} 
\citet{su12093885} &
   &
  X &
   &
   &
   &
   &
   &
   &
   & X
   & X 
   &
   & X 
  \\ 
   & & & & & & & & &  & &   &  & \\
\citet{Garcia2005} &
   &
  X &
  X &
   &
   &
   &
   & X
   & 
   &X
   &X
   &
   &
   \\ 
   \rowcolor[HTML]{EFEFEF} 
   & & & & & & & & &  & &   &  & \\
   \rowcolor[HTML]{EFEFEF} 
\citet{Myers2008ExpressiveAE} &
   &
   &
   X &
   &
   &
   &
   &
   &
   &
   &
   &
  \multicolumn{1}{l}{} &
   \\ 
    & & & & & & & & &  & &   &  & \\
\citet{DBLP:conf/isse/Mont04} &
  X &
   &
  \multicolumn{1}{l}{X} &
   &
   &
   &
   &
   &X
   &
   &
   &
   &X \\ 
   \rowcolor[HTML]{EFEFEF} 
   & & & & & & & & &  & &   &  & \\
    \rowcolor[HTML]{EFEFEF} 
\citet{Jung2014} &
   &
   &
  &
  X &
   &
   &
   &
   &
   &
   &
   &
   &
  \multicolumn{1}{l}{} \\ 
   & & & & & & & & &  & &   &  & \\
\citet{6258293} &
   &
   &
  \multicolumn{1}{l}{} &
  \multicolumn{1}{l}{X} &
  \multicolumn{1}{l}{} &
   &
   &
   &
   &
   &
   &
   &
  \multicolumn{1}{l}{} \\ 
   \rowcolor[HTML]{EFEFEF} 
   & & & & & & & & & &  &   &  & \\
    \rowcolor[HTML]{EFEFEF} 
\citet{CAO2020998} &
  X &
   &
  &
  X &
  &
   &
   & X
   &
   &
   &
   &
   & X \\ 
   & & & & & & & & &  & &   &  & \\
\citet{Schtte2018LUCONDF} &
   &
   &
  \multicolumn{1}{l}{} &
  \multicolumn{1}{l}{X} &
  \multicolumn{1}{l}{} &
   &
   & X
   &
   &
   &
   &
   &
   \\ 
    \rowcolor[HTML]{EFEFEF} 
    & & & & & & & & & & &   &  & \\
     \rowcolor[HTML]{EFEFEF} 
\citet{10.1016/j.jnca.2011.03.019} &
  X &
  X &
  \multicolumn{1}{l}{X} &
  \multicolumn{1}{l}{} &
  \multicolumn{1}{l}{} &
   &
   &
   &
   &
   &
   &
   &
   \\ 
    & & & & & & & & & &  &   &  & \\
\citet{9314823} &
  X &
   &
   &
   &
   &
   &
   &
   &
   &
   &
   &
   &X
   \\ 
    \rowcolor[HTML]{EFEFEF} 
    & & & & & & & & &  & &   &  & \\
     \rowcolor[HTML]{EFEFEF} 
\citet{DBLP:conf/mum/BexhetiL15} &
   &
   &
   &
   & X
   &
   &
   &
   &
   &
   &
   &X
   &X
   \\ 
   & & & & & & & & &  & &   &  & \\
\citet{Rath2013AccessAU} &
  \multicolumn{1}{l}{} &
   &
   &
   &
   &
   &
   &
   &X
   &X
   &X 
   &X 
   &
   \\ 
    \rowcolor[HTML]{EFEFEF} 
    & & & & & & & & &  & &   &  & \\
     \rowcolor[HTML]{EFEFEF} 
\citet{Zrenner2019} &
  \multicolumn{1}{l}{} &
   &
   &
   &
   &
   &
   &
   &
   &X 
   &X
   &
   &X
  \\ 
   & & & & & & & & & &  &   &  & \\
\citet{9119565} &
  \multicolumn{1}{l}{} &
   &
   &
   &
   &
   &
   &
   &
   &
   &X
   &
   &X \\ 
  \rowcolor[HTML]{EFEFEF} 
   & & & & & & & & &  & &   &  & \\
    \rowcolor[HTML]{EFEFEF} 
\citet{DBLP:journals/toit/KeromytisS07} &
  \multicolumn{1}{l}{} &
   &
   &
   &
   &
   &
   &
   &
   &
   &X
   &
   &
   \\ 
    & & & & & & & & &  & &   &  & \\
\citet{hilty2005obligations} &
  \multicolumn{1}{l}{} &
   &
   &
   &
   &
   &
   &
   &
   &
   &
   &
   & X
   \\ 
  \rowcolor[HTML]{EFEFEF} 
   & & & & & & & & &  & &   &  & \\
    \rowcolor[HTML]{EFEFEF} 
\citet{10.1007/978-0-387-73655-6_29} &
  \multicolumn{1}{l}{} &
  \multicolumn{1}{l}{} &
   &
   &
   &
   &
   &
   &
   &
   &
   &
   &X 
   \\ 
   & & & & & & & & & &  &   &  & \\
\citet{DBLP:conf/codaspy/KumariPPK11} &
  \multicolumn{1}{l}{} &
  \multicolumn{1}{l}{} &
   &
   & X
   &
   &
   &
   &
   &
   &
   &
   &
  X \\ 
  \rowcolor[HTML]{EFEFEF} 
   & & & & & & & & & &   &   &  & \\
    \rowcolor[HTML]{EFEFEF} 
\citet{DBLP:conf/icissp/HosseinzadehEJ20} &
 
  \multicolumn{1}{l}{} &
  \multicolumn{1}{l}{} &
   &
   &
   &
   &
   &
   &
   &
   &
   &
   &
  X \\ 
   & & & & & & & & &  & &   &  & \\
\citet{DBLP:conf/sp/Bier13} &
  \multicolumn{1}{l}{} &
  \multicolumn{1}{l}{} &
   &
   &
   &
   &
   &
   &
   &
   &
   &
   &
  X \\ 
   \rowcolor[HTML]{EFEFEF} 
  & & & & & & & & &  & &   &  & \\
   \rowcolor[HTML]{EFEFEF} 
\citet{Lazouski2010UsageCI} &
  \multicolumn{1}{l}{} &
  \multicolumn{1}{l}{} &
   &
   &
   &
   & X
   &
   &
   &
   &
   &
   & \\ 
  & & & & & & & & & &  &   &  & \\
\citet{10.1145/984334.984339} &
  \multicolumn{1}{l}{} &
  \multicolumn{1}{l}{} &
   &
   &
   &
   & X
   &
   &
   &
   &
   &
   & \\ 
    \rowcolor[HTML]{EFEFEF} 
  & & & & & & & & &  & &   &  & \\
   \rowcolor[HTML]{EFEFEF} 
\citet{DBLP:conf/policy/KagalFJ03} &
  \multicolumn{1}{l}{} &
  \multicolumn{1}{l}{} &
   &
   &
   &
   &
   & X
   &
   &
   &
   &
   & \\ 
  
\end{tabular}%
}
\label{tab:requirements}
\end{table}

\subsection{Taxonomy Generation}
The method for taxonomy development proposed by \citet{doi:10.1057/ejis.2012.26} was subsequently used to build a usage control requirements taxonomy. We started by determining the meta-characteristic of our taxonomy (essentially the goal behind its development). In our case, the desire to identify requirements that are used to guide the development of usage control solutions. Following on from this, we identified the condition that would be used to end the taxonomy construction process. In our case, when no new requirements are introduced.

The iterative taxonomy generation method identifies common characteristics that are logical consequences of a meta-characteristic. In our case, the (usage control) specification requirement encapsulates other fine-grained requirements related to usage control specification, such as the \emph{expressiveness} of the usage control policy language. Consequently, individual requirements were grouped according to more general requirements, for example, placing \emph{expressiveness} under \emph{specification}. 
The taxonomy building process ended once the final condition was triggered, i.e., when no further requirements were introduced.

\subsection{Comparison and Synthesis}
The resulting taxonomy was subsequently used to compare the various usage control frameworks that have been proposed to date. The analysis, which was initially performed using Excel spreadsheets, was later synthesized using high level comparative tables and supporting textual descriptions. Finally, building upon the insights gained from our detailed analysis, we derived opportunities and challenges for the usage control domain in general and decentralized systems in particular.

\section{A Taxonomy of Usage Control Requirements} 
\begin{figure}[!t]
  \centering
  \includegraphics[scale=.4]{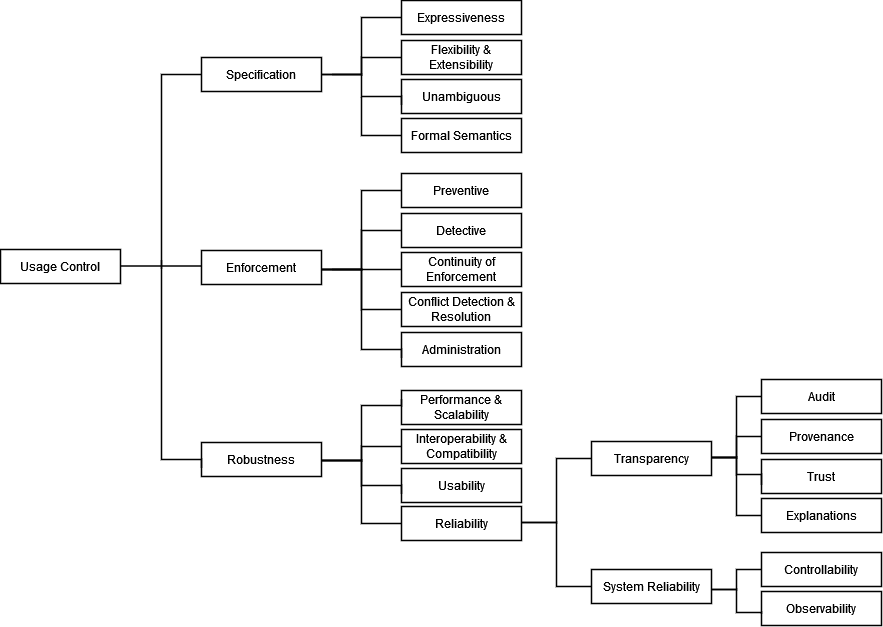}
  \caption{Taxonomy of Usage Control Requirements}
  \label{fig:taxonomy-requirements}
\end{figure}

In the following, we describe our taxonomy of usage control requirements. Table \ref{tab:requirements} displays a matrix of requirements and the corresponding sources from which they are taken. The respective requirements are illustrated in Figure \ref{fig:taxonomy-requirements}, which depicts our final taxonomy. The taxonomy is divided into three high level usage control dimensions: (i) the policy language; (ii) the enforcement mechanism; and (iii) the robustness of the overall solution.

\subsection{Specification}
The specification dimension includes four sub-dimensions that represent requirements relating to policy specification and policy representation. 
\FloatBarrier
\subsubsection{Expressiveness  \cite{Kattinproceedings2008,DBLP:conf/icsnc/HiltyPSSW06,Kumari2010RequirementsAF,DBLP:journals/ieeesp/PretschnerHSSW08,CAO2020998,DBLP:conf/isse/Mont04,pretschner2006distributed,10.1016/j.jnca.2011.03.019,9314823,Zhang2008TowardAU,10.1007/978-3-642-16161-2_19,10.1145/2851613.2851797,6258293,6571694}.} 
According to \citet{pretschner2006distributed}, policy based usage control systems rely on the ability of the policy language to formally express usage control policies by translating high-level policies  defined by  data providers into machine-readable usage policies. 
\citet {DBLP:conf/isse/Mont04} 
argue that the language used to encode usage control policies must be expressive enough such that requests for access to objects can be permitted or prohibited. According to \citet{Zhang2008TowardAU}, various usage control decision components (i.e., obligations, attributes, and conditions) must be encoded using an appropriate language. Moreover, \citet{10.1007/978-1-4419-6794-7_11} introduced a \emph{mutability} decision property, which they argue is an important usage control requirement, as attributes often need to be changed as a side effect of the subject's use of an object.
\citet{CAO2020998,Kumari2010RequirementsAF,DBLP:journals/ieeesp/PretschnerHSSW08}, and \citet{Zhang2008TowardAU} describe different types of conditions, which relate to time, cardinality, purpose of use, and technical or governance constraints, which may need to be included in usage control policies, in order to accommodate different application contexts (e.g., privacy, copyright protection, regulations) and different fields (e.g., IoT and \emph{information and communications technology} (ICT)). 
Moreover, several studies \cite{DBLP:conf/icsnc/HiltyPSSW06,10.1007/978-3-642-16161-2_19,10.1145/2851613.2851797,6258293,6571694} highlight the need for policies that can specify rules with respect to contextual information, such as GPS information in the context of mobile and ubiquitous applications.

\subsubsection{Flexibility \& Extensibility \cite{su12093885,Garcia2005,10.1016/j.jnca.2011.03.019}.} Usage control can be applied in different application contexts (e.g., \emph{digital rights management} (DRM) and data privacy) or different fields (e.g., ICT and IoT). This, as highlighted by \citet {su12093885}, gives rise to the need for flexible usage control solutions that can be adapted to cater for various use case scenarios. Besides, \citet{Garcia2005} and \citet{10.1016/j.jnca.2011.03.019} highlight the need for an extensible usage control solution that allows for new types of policies to be supported at a later point in time.

\subsubsection{Unambiguous \cite{Myers2008ExpressiveAE,DBLP:conf/isse/Mont04,Garcia2005,10.1016/j.jnca.2011.03.019}.} In order to automatically enforce usage control policies, said policies need to be translated from high-level goals to formal rules, such that they can be deployed and enforced by the usage control system. This is only possible if the usage control policy language is able to unambiguously specify the meaning of such policies via a well-defined syntax and semantics \cite{DBLP:conf/isse/Mont04,Garcia2005}. According to \citet{Myers2008ExpressiveAE} and \citet{10.1016/j.jnca.2011.03.019}, expressive, formal, and well-defined information policies can ensure the correct enforcement of high level goals.

\subsubsection{Formal Semantics \cite{Jung2014,6258293,CAO2020998,Schtte2018LUCONDF}.} Formal semantics refers to approaches that are used to specify the precise meaning for the various concepts and rules encoded in usage control policies. According to \citet{Lazouski2010UsageCI} and \citet{Schtte2018LUCONDF}, the formalization of usage control policies helps to facilitate system governance by verifying compliance against higher level goals. Moreover, \citet{Jung2014,6258293} and \citet{CAO2020998} state that formal policies can help with automated analysis (i.e., automatizing the decision of the system and checking for policy conflicts). 

\subsection{Enforcement}
Enforcement refers to the mechanisms used to enforce and manage usage policies throughout the usage process, which consists of three phases: before usage, ongoing usage, and after usage \cite {9314823}. According to \citet{pretschner2006distributed}, in order for usage control policies to work as intended, the policy enforcement must be applied continuously. 

\subsubsection{Preventive \cite{DBLP:journals/ieeesp/PretschnerHSSW08,DBLP:conf/codaspy/KumariPPK11,DBLP:conf/mum/BexhetiL15,10.1145/984334.984339}.} According to \citet{DBLP:journals/ieeesp/PretschnerHSSW08,DBLP:conf/codaspy/KumariPPK11,DBLP:conf/mum/BexhetiL15}, and \citet{10.1145/984334.984339}, the dynamic and proactive enforcement of data usage policies implies the ability of the usage control solution, in particular, the preventive mechanism to at least: \inlinerom{\item allow or prohibit requests for data usage; \item revoke access in the event of policy violations; \item delay an attempted usage request until the corresponding obligations are fulfilled; \item update user or object attributes as a result of usage decisions \item execute actions such as sending notifications to data owners.} 

\subsubsection{Detective \cite{pretschner2006distributed}.} According to \citet{pretschner2006distributed}, detective mechanisms are very important, particularly if the usage control framework is not able to dynamically enforce the policy restrictions or prevent policy violations from happening. For instance, it is difficult to see if data are actually deleted, but there may be technical means to show that the respective command has been executed by using  different detective mechanisms, such as auditing, logging, or simply notifying a user when the command is executed. 

\subsubsection{Continuity of Enforcement \cite{Lazouski2010UsageCI,10.1145/984334.984339,pretschner2006distributed}.} According to \citet{Lazouski2010UsageCI,10.1145/984334.984339}, and \citet{pretschner2006distributed}, a usage control framework should be capable of handling the continuous enforcement of policies. This implies the management of attributes, conditions, and the fulfillment of obligation actions that reflect the validity of the continuous usage of data objects.

\subsubsection{Conflict Detection  \& Resolution
\cite{DBLP:conf/policy/KagalFJ03,Garcia2005,CAO2020998,Schtte2018LUCONDF}.} 
Another aspect of enforcement is the management of conflicting rules, which is particularly difficult in decentralized or distributed systems, as data may be governed by a variety of policies \cite{DBLP:conf/policy/KagalFJ03,Garcia2005}. 
According to \citet{Schtte2018LUCONDF} and \citet{CAO2020998}, an enforcement engine should be able to detect and resolve conflicting or incomplete rules. 

\subsubsection{Administration \cite{Rath2013AccessAU,DBLP:conf/isse/Mont04}.} 
According to \citet{DBLP:conf/isse/Mont04}, a complete usage control solution must include an administration tool, which provides an interface to manage (i.e., create, edit, and delete) usage control policies. 
\citet{Rath2013AccessAU} highlight the importance of administration interfaces, especially for users that use healthcare information systems, as such a feature allows them to customize their usage policies without the help of system administrators.

\subsection{Robustness}
Robustness is an all encompassing term used to refer to requirements that relate to the overall effectiveness of the usage control system. 

\subsubsection{Performance \& Scalability \cite{Rath2013AccessAU,Zrenner2019,su12093885,9119565,Garcia2005,DBLP:journals/toit/KeromytisS07}.}  
    Several works \cite{Zrenner2019,Rath2013AccessAU,su12093885,9119565,Garcia2005} highlight that a usage control infrastructure should be performant enough to cater for the parallel processing of a large number of requests by data requestors and for a short loading time of the requested data.
    Furthermore,
    \citet{DBLP:journals/toit/KeromytisS07} and \citet{Zrenner2019} highlight the fact that policy management systems need to be realized using scalable architectures that can handle an increasing number of users.
    \subsubsection{Interoperability \& Compatibility \cite{Rath2013AccessAU,Zrenner2019,su12093885,Garcia2005,DBLP:journals/ieeesp/PretschnerHSSW08}.} \citet{Zrenner2019} state that usage control needs to work even if data providers and data consumers have different infrastructures. While, \citet{su12093885,Rath2013AccessAU}, and \citet{Garcia2005} highlight the importance of establishing interoperable mechanisms as without them the system can only be utilized on particular devices, which according to \citet{DBLP:journals/ieeesp/PretschnerHSSW08} will limit the uptake of usage control technologies.
    Besides, \citet{DBLP:conf/drm/JamkhedkarHL10} attest that it is necessary to have mechanisms that allow for the specification of usage policies that can be interpreted and enforced across multiple different computing environments. This can be fulfilled by: (i) applying standards; and (ii) separating the usage control system components for policy expression, policy interpretation, and policy enforcement. Such separation allows policies to express the restrictions with minimal apriori knowledge of the IT environments in which the policies will be interpreted. 
    \subsubsection{Usability  
    \cite{DBLP:conf/icsnc/HiltyPSSW06, DBLP:conf/mum/BexhetiL15,Rath2013AccessAU,Kumari2010RequirementsAF}.} According to \citet{Rath2013AccessAU} and \citet{DBLP:conf/icsnc/HiltyPSSW06}, the usability of administration tools is an important requirement that should be supported via user-friendly interfaces or easy-to-use languages.
    \citet{DBLP:conf/icsnc/HiltyPSSW06} state that administration tools may be used by end users (as opposed to system administrators) in order to enter their preferences and manage the use of their data. 
   Usability is especially important when it comes to  mobile and/or web applications \cite{DBLP:conf/mum/BexhetiL15,Rath2013AccessAU,Kumari2010RequirementsAF} where users need to manage  how their data are used.
    \subsubsection{Reliability \cite{DBLP:journals/ieeesp/PretschnerHSSW08,hilty2005obligations,DBLP:conf/isse/Mont04,9119565,10.1007/978-0-387-73655-6_29,DBLP:conf/mum/BexhetiL15,DBLP:conf/codaspy/KumariPPK11,DBLP:conf/icissp/HosseinzadehEJ20,Kumari2010RequirementsAF,Zrenner2019,CAO2020998,DBLP:conf/sp/Bier13,su12093885,pretschner2006distributed}.} Reliability refers to the ability of a usage control mechanism to be compliant with usage control policies and transparent with respect to the way data are used \cite{pretschner2006distributed}. 
    %
    \paragraph{System Reliability
    \cite{DBLP:journals/ieeesp/PretschnerHSSW08,hilty2005obligations,DBLP:conf/isse/Mont04,9119565,10.1007/978-0-387-73655-6_29}.}
    \citet{DBLP:journals/ieeesp/PretschnerHSSW08,DBLP:conf/isse/Mont04}, and \citet{9119565} highlight the importance of system reliability when dealing with distributed usage control systems, as sometimes the infrastructure of the provider is hosted by a third party, leaving the owner with little or no control over how their data are used.
    System reliability depends on the level of compliance with respect to usage control policies that can be ensured by two factors: controllability and observability, notions initially introduced by \citet{hilty2005obligations}. 
    On one hand, controllable restrictions are policies by which the data provider can ensure that the data consumer complies with everything that has been mentioned in the policy. On the other hand, observable restrictions are policies that the data provider cannot control. In this context, the data provider can only observe violations and take compensatory actions, such as lowering the trust or credibility rating of the data consumer or taking some form of legal action.
    \paragraph{Transparency \cite{DBLP:conf/mum/BexhetiL15,DBLP:conf/codaspy/KumariPPK11,DBLP:conf/icissp/HosseinzadehEJ20,Kumari2010RequirementsAF,Zrenner2019,CAO2020998,DBLP:conf/sp/Bier13,su12093885,pretschner2006distributed,9314823}.} The usage control process should be transparent to and comprehensible by data providers and data consumers alike \cite{DBLP:conf/mum/BexhetiL15,DBLP:conf/codaspy/KumariPPK11,Zrenner2019}. 
    \citet{pretschner2006distributed} mention data auditing as a means to observe the fulfillment or the violation of non-observable restrictions.
    Moreover, \citet{DBLP:conf/icissp/HosseinzadehEJ20} state that the logging of data usage information can ensure both transparency and reliability, as logs provide visibility with respect to the inner workings of a software systems.
    \citet{DBLP:conf/sp/Bier13} and \citet{su12093885} indicate that data provenance tracking is complementary to distributed data usage monitoring. 
    \citet{CAO2020998} and \citet{Kumari2010RequirementsAF} state that providing explanations, with respect to actions performed by the system, is part of ensuring transparency as explanations provide a basis for decision-making. 
    While, \citet{su12093885}, \citet{CAO2020998}, and \citet{9314823} highlight the importance of a trusted infrastructure, especially when it comes to guaranteeing secure data sharing and adherence to usage policies.

\afterpage{%
\begingroup
\tiny
\centering
\begin{longtable}[c]{p{2cm}lp{2.5cm}p{2.1cm}ll}
\caption{High Level Overview of Existing Usage Control Frameworks}
\label{tab:general-description-v1}\\

\rowcolor[HTML]{9B9B9B} 
\cellcolor[HTML]{9B9B9B} &
  \cellcolor[HTML]{9B9B9B} &
  \cellcolor[HTML]{9B9B9B} &
  \cellcolor[HTML]{9B9B9B} &
  \cellcolor[HTML]{9B9B9B} &
  \cellcolor[HTML]{9B9B9B} \\
\rowcolor[HTML]{9B9B9B} 
\multirow{-2}{*}{\cellcolor[HTML]{9B9B9B}\textbf{Framework}} &
  \multirow{-2}{*}{\cellcolor[HTML]{9B9B9B}\textbf{Name}} &
  \multirow{-2}{*}{\cellcolor[HTML]{9B9B9B}\textbf{Domain}} &
  \multirow{-2}{*}{\cellcolor[HTML]{9B9B9B}\textbf{Specification}} &
  \multirow{-2}{*}{\cellcolor[HTML]{9B9B9B}\textbf{Enforcement}} &
  \multirow{-2}{*}{\cellcolor[HTML]{9B9B9B}\textbf{Robustness}} \\ 
\endfirsthead
\multicolumn{6}{c}%
{{\bfseries Table \thetable\ continued from previous page}} \\

\rowcolor[HTML]{9B9B9B} 
\cellcolor[HTML]{9B9B9B} &
  \cellcolor[HTML]{9B9B9B} &
  \cellcolor[HTML]{9B9B9B} &
  \cellcolor[HTML]{9B9B9B} &
  \cellcolor[HTML]{9B9B9B} &
  \cellcolor[HTML]{9B9B9B} \\
\rowcolor[HTML]{9B9B9B} 
\multirow{-2}{*}{\cellcolor[HTML]{9B9B9B}\textbf{Framework}} &
  \multirow{-2}{*}{\cellcolor[HTML]{9B9B9B}\textbf{Name}} &
  \multirow{-2}{*}{\cellcolor[HTML]{9B9B9B}\textbf{Domain}} &
  \multirow{-2}{*}{\cellcolor[HTML]{9B9B9B}\textbf{Specification}} &
  \multirow{-2}{*}{\cellcolor[HTML]{9B9B9B}\textbf{Enforcement}} &
  \multirow{-2}{*}{\cellcolor[HTML]{9B9B9B}\textbf{Robustness}} \\ 
\endhead

\citet{10.1007/978-3-642-16161-2_19} &
  ConUCON &
  mobile, cloud, IoT, and industry 4.0 &
  \begin{tabular}[c]{@{}l@{}}obligations, conditions, \\ mutable attributes,\\ context\end{tabular} &
  \begin{tabular}[c]{@{}l@{}}continuity,\\ preventive,\\ administration\end{tabular} &
  \begin{tabular}[c]{@{}l@{}}implementation,\\ performance evaluation\end{tabular} \\ 
  
\rowcolor[HTML]{EFEFEF}   
  \citet{Baldini2013,Neisse2015SecKitAM} &
  \begin{tabular}[c]{@{}l@{}}trusted usage \\ framework\end{tabular} &
  mobile, cloud, IoT, and industry 4.0 &
  \begin{tabular}[c]{@{}l@{}}logic-based\\ \\ obligations, conditions,\\ attributes, context\end{tabular} &
  \begin{tabular}[c]{@{}l@{}}continuity,\\preventive,\\ detective,\\ conflicts resolution,\\ administration\end{tabular} &
  \begin{tabular}[c]{@{}l@{}}implementation,\\ performance evaluation,\\ use case validation\end{tabular} \\ 
  
\citet{CAO2020998} &
  \begin{tabular}[c]{@{}l@{}}trustworthy data\\ sharing platform\end{tabular} &
  mobile, cloud, IoT, and industry 4.0 &
  \begin{tabular}[c]{@{}l@{}}logic-based\\ \\ obligations, conditions, \\attributes, context\end{tabular} &
  \begin{tabular}[c]{@{}l@{}}continuity,\\preventive,\\ administration,\\ conflicts resolution\end{tabular} &
  \begin{tabular}[c]{@{}l@{}}implementation,\\ performance evaluation,\\ use case validation\end{tabular} \\ 
 
\rowcolor[HTML]{EFEFEF} 
 \citet{CARNIANI201637}, \citet{6470943} &
  \begin{tabular}[c]{@{}l@{}}usage control \\ systems - proposal\end{tabular} &
  mobile, cloud, IoT, and industry 4.0 &
  \begin{tabular}[c]{@{}l@{}}obligations, conditions,  \\ mutable attributes,\\ context\end{tabular} &
  \begin{tabular}[c]{@{}l@{}}continuity,\\preventive,\\ administration\end{tabular} &
  \begin{tabular}[c]{@{}l@{}}implementation,\\ performance evaluation\end{tabular} \\ 
 
 \citet{8456147} &
  \begin{tabular}[c]{@{}l@{}}usage control \\ systems - extension\end{tabular} &
  mobile, cloud, IoT, and industry 4.0 &
  \begin{tabular}[c]{@{}l@{}}obligations, conditions, \\ mutable attributes,\\ context\end{tabular} &
  \begin{tabular}[c]{@{}l@{}}continuity,\\preventive, \\ administration\end{tabular} &
  \begin{tabular}[c]{@{}l@{}}implementation,\\ performance evaluation, \\ use case validation\end{tabular} \\ 

\rowcolor[HTML]{EFEFEF} 
\citet{9314823} &
  IntentKeeper &
  mobile, cloud, IoT, and industry 4.0 &
  obligations, conditions &
  \begin{tabular}[c]{@{}l@{}}preventive,   \\ administration\end{tabular} &
  \begin{tabular}[c]{@{}l@{}}implementation,\\ performance evaluation,\\ use case validation\end{tabular} \\  
 
\citet{6258293} &
  \_ &
  mobile, cloud, IoT, and industry 4.0 &
  \begin{tabular}[c]{@{}l@{}}logic-based\\ \\ obligations, conditions\end{tabular} &
  \begin{tabular}[c]{@{}l@{}}continuity,\\preventive,\\ detective,\\ administration\end{tabular} &
  \begin{tabular}[c]{@{}l@{}}implementation,\\ performance evaluation\end{tabular} \\ 

\rowcolor[HTML]{EFEFEF} 
\citet{10.1007/978-3-319-68063-7_8} &
  \begin{tabular}[c]{@{}l@{}}usage control \\ systems - extension\end{tabular} &
  mobile, cloud, IoT, and industry 4.0 &
  \begin{tabular}[c]{@{}l@{}}obligations, conditions, \\ mutable attributes,\\ context\end{tabular} &
  \begin{tabular}[c]{@{}l@{}}continuity,\\preventive, \\ administration\end{tabular} &
  \begin{tabular}[c]{@{}l@{}}implementation,\\ performance evaluation, \\ use case validation\end{tabular} \\ 
  
\citet{Jung2014} &
  IND\textsuperscript{2}UCE &
  \begin{tabular}[c]{@{}l@{}}mobile, cloud, IoT, and\\ industry 4.0\\ \\ domain-agnostic\end{tabular} &
  \begin{tabular}[c]{@{}l@{}}logic-based\\ \\ obligations, conditions,
  \\attributes, context\end{tabular} &
  \begin{tabular}[c]{@{}l@{}}continuity,\\preventive,\\ detective,\\ administration,\end{tabular} &
  implementation \\ 

\rowcolor[HTML]{EFEFEF} 
\citet{kateb2014} &
  OB-XACML &
  domain-agnostic &
  obligations, attributes, context &
  \begin{tabular}[c]{@{}l@{}}continuity,\\preventive\end{tabular} &
  implementation \\ 
 
\citet{Lazouski2012APF} &
  U-XACML &
  mobile, cloud, IoT, and industry 4.0 &
  \begin{tabular}[c]{@{}l@{}}obligations, conditions,\\ mutable attributes,\\ context\end{tabular} &
  \begin{tabular}[c]{@{}l@{}}continuity,\\preventive,\\ administration\end{tabular} &
  \begin{tabular}[c]{@{}l@{}}implementation,\\ performance evaluation\end{tabular} \\ 

\rowcolor[HTML]{EFEFEF} 
\citet{9161997,8717790} &
  \begin{tabular}[c]{@{}l@{}}usage control \\ systems - extension\end{tabular} &
  \begin{tabular}[c]{@{}l@{}}networking, \\ operating systems\\ and collaborative software\end{tabular} &
  \begin{tabular}[c]{@{}l@{}}obligations, conditions, \\ mutable attributes,\\ context\end{tabular} &
  \begin{tabular}[c]{@{}l@{}}continuity,\\preventive,\\ administration\end{tabular} &
  \begin{tabular}[c]{@{}l@{}}implementation,\\ performance evaluation,\\ use case validation\end{tabular} \\

\citet{8029555} &
  \begin{tabular}[c]{@{}l@{}}usage control \\ systems - extension\end{tabular} &
  mobile, cloud, IoT, and industry 4.0 &
  \begin{tabular}[c]{@{}l@{}}obligations, conditions, \\ mutable attributes,\\ context\end{tabular} &
  \begin{tabular}[c]{@{}l@{}}continuity,\\preventive,\\ administration\end{tabular} &
  \begin{tabular}[c]{@{}l@{}}implementation,\\ performance evaluation,\\ use case validation\end{tabular} \\ 
  
\rowcolor[HTML]{EFEFEF}   
\citet{Marra2019} &
  \begin{tabular}[c]{@{}l@{}}usage control\\ systems - extension\end{tabular} &
  mobile, cloud, IoT, and industry 4.0 &
  \begin{tabular}[c]{@{}l@{}}obligations, conditions, \\ mutable attributes,\\ context\end{tabular} &
  \begin{tabular}[c]{@{}l@{}}continuity,\\preventive, \\ administration\end{tabular} &
  \begin{tabular}[c]{@{}l@{}}implementation,\\ performance evaluation, \\ use case validation\end{tabular} \\

\citet{10.1007/978-3-319-72389-1_43} &
  \begin{tabular}[c]{@{}l@{}}usage control\\ systems - extension\end{tabular} &
  \begin{tabular}[c]{@{}l@{}}networking, \\ operating systems\\ and collaborative software\end{tabular} &
  \begin{tabular}[c]{@{}l@{}}obligations, conditions, \\ mutable attributes,\\ context\end{tabular} &
  \begin{tabular}[c]{@{}l@{}}continuity,\\preventive, \\ administration\end{tabular} &
  \begin{tabular}[c]{@{}l@{}}implementation,\\ performance evaluation, \\ use case validation\end{tabular} \\ 

\rowcolor[HTML]{EFEFEF} 
\citet{Martinelli2019ObligationMI} &
  \begin{tabular}[c]{@{}l@{}}usage control\\ systems - extension\end{tabular} &
  mobile, cloud, IoT, and industry 4.0 &
  \begin{tabular}[c]{@{}l@{}}obligations, conditions, \\ mutable attributes,\\ context\end{tabular} &
  \begin{tabular}[c]{@{}l@{}}continuity,\\preventive, \\ administration\end{tabular} &
  \_ \\ 
  
\citet{10.1145/2851613.2851797} &
  \begin{tabular}[c]{@{}l@{}}usage control\\ systems - extension\end{tabular} &
  mobile, cloud, IoT, and industry 4.0 &
  \begin{tabular}[c]{@{}l@{}}obligations, conditions, \\ mutable attributes,\\ context\end{tabular} &
  \begin{tabular}[c]{@{}l@{}}continuity,\\preventive, \\ administration\end{tabular} &
  \begin{tabular}[c]{@{}l@{}}implementation,\\ use case validation\end{tabular} \\ 
  
\rowcolor[HTML]{EFEFEF} 
\citet{su12093885,MunozArcentales2019AnAF} &
  \_ &
  mobile, cloud, IoT, and industry 4.0 &
  obligations, conditions &
  \begin{tabular}[c]{@{}l@{}}continuity,\\preventive, \\ administration\end{tabular} &
  \begin{tabular}[c]{@{}l@{}}implementation,\\ performance evaluation,\\ use case validation\end{tabular} \\ 

\citet{6045968} &
  \_ &
  domain-gnostic &
  \begin{tabular}[c]{@{}l@{}}logic-based\\ \\ obligations, conditions\end{tabular} &
  \begin{tabular}[c]{@{}l@{}}continuity,\\preventive,\\ detective,\\ administration\end{tabular} &
  \begin{tabular}[c]{@{}l@{}}implementation,\\ performance evaluation\end{tabular} \\ 
 
\rowcolor[HTML]{EFEFEF}   
\citet{5319024} &
  xDUCON &
  \begin{tabular}[c]{@{}l@{}}networking, \\ operating systems and  \\ collaborative software\end{tabular} &
  \begin{tabular}[c]{@{}l@{}}obligations, conditions,\\ mutable attributes,\\ context\end{tabular} &
  \begin{tabular}[c]{@{}l@{}}continuity,\\preventive\end{tabular} &
  \_ \\ 
  
\citet{Schtte2018LUCONDF} &
  LUCON &
  \begin{tabular}[c]{@{}l@{}}networking, \\ operating Systems\\ and collaborative Software\end{tabular} &
  \begin{tabular}[c]{@{}l@{}}logic-based\\ \\ obligations\end{tabular} &
  \begin{tabular}[c]{@{}l@{}}continuity,\\preventive,\\ detective,\\ conflicts resolution,\\ administration\end{tabular} &
  \begin{tabular}[c]{@{}l@{}}implementation,\\ performance evaluation\end{tabular} \\ 

\rowcolor[HTML]{EFEFEF} 
\citet{10.1016/j.jnca.2011.03.019} &
  \_ &
  \begin{tabular}[c]{@{}l@{}}networking, \\ operating systems and  \\ collaborative software\end{tabular} &
  \begin{tabular}[c]{@{}l@{}}obligations, conditions,\\ mutable attributes\end{tabular} &
   \begin{tabular}[c]{@{}l@{}}continuity,\\preventive \end{tabular} &
  \begin{tabular}[c]{@{}l@{}}implementation,\\ performance evaluation\end{tabular} \\

\citet{Silva2010} &
  \_ &
  \begin{tabular}[c]{@{}l@{}}networking, \\ operating systems   \\ and collaborative software\\ \\ domain-agnostic\end{tabular} &
  \begin{tabular}[c]{@{}l@{}}logic-based\\ \\ obligations, conditions\end{tabular} &
  \begin{tabular}[c]{@{}l@{}}continuity,\\preventive, \\ detective,  \\ administration\end{tabular} &
  \begin{tabular}[c]{@{}l@{}}implementation,\\ performance evaluation\end{tabular} \\ 
  
\rowcolor[HTML]{EFEFEF} 
\citet{Wchner2013CompliancePreservingCS} &
  \_ &
  mobile, cloud, IoT, and industry 4.0 &
  \begin{tabular}[c]{@{}l@{}}logic-based\\ \\ obligations, conditions\end{tabular} &
  \begin{tabular}[c]{@{}l@{}}continuity,\\preventive,\\ detective,\\ administration\end{tabular} &
  \begin{tabular}[c]{@{}l@{}}implementation,\\ performance evaluation\end{tabular} \\ 
  
\citet{6405363} &
  \_ &
  \begin{tabular}[c]{@{}l@{}}networking, \\ operating systems and\\ collaborative software\\ \\ domain-agnostic\end{tabular} &
  \begin{tabular}[c]{@{}l@{}}logic-based\\ \\ obligations, conditions\end{tabular} &
  \begin{tabular}[c]{@{}l@{}}continuity,\\preventive,\\ detective,\\ administration\end{tabular} &
  \begin{tabular}[c]{@{}l@{}}implementation,\\ performance evaluation\end{tabular} \\ 

\rowcolor[HTML]{EFEFEF} 
\citet{Xu2007} &
  \_ &
  \begin{tabular}[c]{@{}l@{}}networking, \\ operating systems and  \\ collaborative software\end{tabular} &
  mutable attributes &
  \begin{tabular}[c]{@{}l@{}}continuity,\\preventive,\\ administration\end{tabular} &
  \begin{tabular}[c]{@{}l@{}}implementation,\\ performance evaluation\end{tabular} \\ 
 
\citet{Zhang2008TowardAU} &
  \_ &
  \begin{tabular}[c]{@{}l@{}}networking, \\ operating systems \\ and collaborative software\end{tabular} &
  \begin{tabular}[c]{@{}l@{}}obligations, conditions, \\ mutable attributes,\\ context\end{tabular} &
   \begin{tabular}[c]{@{}l@{}}continuity,\\preventive\end{tabular} &
  \begin{tabular}[c]{@{}l@{}}implementation,\\ performance evaluation\end{tabular} \\ 

\end{longtable}
\endgroup
}%
\raggedbottom

\section{Frameworks}
The goal of this section is to provide a detailed analysis of existing usage control frameworks with a particular focus on the domain of usage, as well as their support for the various policy specification, enforcement, and robustness requirements. 

\subsection{Domain of Usage}
In Table \ref{tab:general-description-v1} we provide a high level overview of the domain of usage of the various usage control frameworks. The specification, enforcement, and robustness columns of the table, which serve to provide a single snapshot of existing usage control proposals, are discussed in detail in the subsequent sub-sections.
Due to the variety of usage control domains,  we group together the usage control framework proposals according to three application domains, namely,
\emph {mobile, cloud, IoT, and industry 4.0};
\emph {networking, operating systems and collaborative software}; and \emph {domain-agnostic}. 

\subsubsection {Mobile, Cloud, IoT, and Industry 4.0.} 
A prominent framework found in the literature is the \emph{usage control systems} framework proposed by \citet{6470943} and \citet{CARNIANI201637},
which can be used to control data usage in modern decentralized and distributed environments, for instance, IoT, cloud computing, mobile computing, and data sharing platforms. 
The same framework was later refined and used in various industry 4.0 use cases \cite{10.1007/978-3-319-68063-7_8,8456147,8029555,Marra2019}. Additionally, an adaptation of the proposed framework is used by \citet{10.1145/2851613.2851797} in a mobile computing context.
The \emph{intent-oriented data usage control for federated data analytics} or IntentKeeper framework proposed by \citet{9314823}, which protects data during federated data analytics, is applied in an automotive setting. 
The \emph{trusted usage framework} proposed by \citet{Baldini2013,Neisse2015SecKitAM} is intended to  address the challenges of heterogeneity in  IoT technologies.
The \emph{trustworthy data sharing platform} framework proposed by \citet{CAO2020998} is designed with a smart city use case in mind. Another framework that deals with data sharing in smart cities is proposed by \citet{MunozArcentales2019AnAF} was later implemented by the same authors \cite{su12093885}.
In turn, the \emph{integrated distributed data usage control enforcement} (IND\textsuperscript{2}UCE) \cite{Jung2014} framework is designed to support usage control in cloud environments. IND\textsuperscript{2}UCE is originally proposed by \citet{Steinebach2016} in order to manage data usage in industry 4.0 environments. A technical implementation of IND\textsuperscript{2}UCE, under the name \emph{MYDATA technologies}, is described in \cite{IND2UCEMAYDATA}.
The framework is also mentioned as a solution for business ecosystems in \cite{Zrenner2019}.
The framework proposed by \citet{Wchner2013CompliancePreservingCS} is designed to preserve regulatory compliance in federated cloud storage.
The \emph{context-aware usage control} (ConUCON) framework \cite{10.1007/978-3-642-16161-2_19} is designed for mobile computing and uses contextual information in order to enhance data protection.
While, the framework proposed by \citet{6258293} aims to enhance Android security by allowing users to manage fine-grained security policies. 

\subsubsection {Networking, Operating Systems and Collaborative Software.}
The \emph{usage control systems} framework proposed by \citet{6470943} and \citet{CARNIANI201637} is further refined and used to enhance network security \cite{9161997,8717790,10.1007/978-3-319-72389-1_43}.
The xDUCON framework, proposed by \citet{5319024} is intended to regulate how data are managed and shared among distributed environments and organizations.
While, LUCON \cite{Schtte2018LUCONDF} is a message-based system that guarantees the secrecy of messages routed between services. LUCON also appears in the context of \emph{MYDATA technologies} for controlling data flows between endpoints.
In turn, the framework presented in \cite{6405363} aims to monitor the use of confidential data at the operating system level.
The frameworks proposed by \citet{10.1016/j.jnca.2011.03.019} and \citet{Xu2007} cater for the management of operating systems resources (e.g., files, network connections, memory areas, and system applications).
While, the framework proposed by \citet{Silva2010} is evaluated in an operating system context.
Finally, the framework proposed by \citet{Zhang2008TowardAU} is applied in a collaborative systems context. An architectural instantiation of the same framework with some improvements can be found in \cite{Kattinproceedings2008}.

\subsubsection {Domain-agnostic.}
The framework presented in \cite{Lazouski2012APF} is designed for modern distributed computing systems, without a specific use case in mind. 
Although the IND\textsuperscript{2}UCE framework \cite{Jung2014} is applied in a cloud computing context, the framework was originally proposed for controlling data usage in modern distributed environments, in general. 
Other usage control frameworks proposed by \citet{Silva2010} and \citet{6405363} that are meant to be domain agnostic are evaluated using operating system use cases. 
While, \citet{Martinelli2019ObligationMI} demonstrate the effectiveness of the \emph{usage control systems} framework proposed by \citet{6470943} and \citet{CARNIANI201637} as a general purpose architecture.
%
The domain-agnostic OB-XACML framework, proposed by
\citet{kateb2014}, deals with enhancing XACML in order to cater for usage control obligations and continuity of enforcement.
In turn, the framework proposed by \citet{6045968} can be used to specify and enforce general purpose usage control policies.

\subsection{Specification}
According to \citet{DBLP:conf/sp/Bier13}, when it comes to policy specification, one has to differentiate between the policy language, the representation format, and the model underpinning the usage control system. 
In Table \ref{tab:specification}, we present a comparison of the various approaches used for policy specification.

\subsubsection{Expressiveness.}
The expressiveness of a policy language is reflected by the different decision factors used to express high-level usage control policies. Decision factors depend on the specification of policy rules, conditions, subject and object attributes, as well as contextual information.

\begin{table}[t!]
\centering
\tiny
\caption[Caption for LOF]{Usage Control Policy Specification \protect\footnotemark}
\begin{tabular}{p{2.1cm}p{1cm}lp{1cm}p{1cm}p{1cm}p{1.2cm}p{1.2cm}p{0.9cm}}
\rowcolor[HTML]{C0C0C0} 
\textbf{Framework} &
  \multicolumn{4}{l}{\cellcolor[HTML]{C0C0C0}\textbf{Expressiveness}} &
  \multicolumn{2}{l}{\cellcolor[HTML]{C0C0C0}\textbf{Flexibility \& Extensibility}} &
  \textbf{Unambiguous} & \textbf{Formal Semantics}\\ 
\rowcolor[HTML]{C0C0C0} 
 &
  \textbf{Operators/ Rules} &
  \textbf{Conditions} &
  \textbf{Attributes} &
  \textbf{Context} &
   \textbf{\begin{tabular}[c]{@{}l@{}}
  Represent-\\ation
  \end{tabular}}&
  \textbf{Model} & &
   \\ 
\citet{10.1007/978-3-642-16161-2_19} &
  A, O &
  \begin{tabular}[c]{@{}l@{}}environmental \\ system-oriented\end{tabular} &
  mutability &
  \begin{tabular}[c]{@{}l@{}}model\\ based\end{tabular} &
  XML &
  UCON &
  \_ & \_\\ 

\rowcolor[HTML]{EFEFEF} 
\citet{Baldini2013,Neisse2015SecKitAM} &E, C, A
   &\begin{tabular}[c]{@{}l@{}}cardinal\\temporal\\event-defined \end{tabular}
   &
   &\begin{tabular}[c]{@{}l@{}}model\\ based \end{tabular}
   &XML
   &\_
   &\_&OSL  
   \\ 

\citet{CAO2020998} &
  P, Pr, O &
  \begin{tabular}[c]{@{}l@{}}actor\\ spatial\\ temporal\\ purpose\\ monetization\end{tabular} &
   &
  \begin{tabular}[c]{@{}l@{}}model\\ based\end{tabular} &
  XML &
  DUPO &
  \_ & defeasible logic \\ 

\rowcolor[HTML]{EFEFEF}
\citet{6470943,CARNIANI201637}
& A, O & \begin{tabular}[c]{@{}l@{}}environmental\\ system-oriented\end{tabular} & mutability &
\begin{tabular}[c]{@{}l@{}}system\\ based\end{tabular} & XML & XACML &\_ & \_\\ 

\citet{8456147}
& A, O & \begin{tabular}[c]{@{}l@{}}environmental\\ system-oriented\end{tabular} & mutability &
\begin{tabular}[c]{@{}l@{}}system\\ based\end{tabular} & XML & XACML &\_ & \_\\ 

\rowcolor[HTML]{EFEFEF}
\citet{9314823} &
  P, Pr, O &
  \begin{tabular}[c]{@{}l@{}}purpose\\ temporal\\ spatial\\ event-defined\end{tabular} &
  \_ &
  \_ &
  XML &
  ODRL &
  \_& \_ \\ 
\citet{6258293}
  &
  E, C, Ac &
  \begin{tabular}[c]{@{}l@{}}cardinal\\ temporal\\ spatial\end{tabular} &
  \_ &
  \_ &
  XML &
  \_ &
  \_ & OSL\\ 
 
\rowcolor[HTML]{EFEFEF}
\citet{10.1007/978-3-319-68063-7_8}
& A, O & \begin{tabular}[c]{@{}l@{}}environmental\\ system-oriented\end{tabular} & mutability &
\begin{tabular}[c]{@{}l@{}}system\\ based\end{tabular} & XML & XACML &\_ & \_\\ 

\citet{Jung2014} &
  E, C, Ac &
  \begin{tabular}[c]{@{}l@{}}cardinal \\ temporal\end{tabular} &
   &
  \begin{tabular}[c]{@{}l@{}}system-\\ based\end{tabular} &
  XML &\_
   &
  \_ 
  & OSL\\ 

\rowcolor[HTML]{EFEFEF}
\citet{kateb2014} &
  A, O &
  \begin{tabular}[c]{@{}l@{}}environmental\\ system oriented\end{tabular} &
   &
  \begin{tabular}[c]{@{}l@{}}model\\ based\end{tabular} &
  XML &
  XACML &
  \_& \_ \\ 
  
\citet{Lazouski2012APF}
& A, O & \begin{tabular}[c]{@{}l@{}}environmental\\ system-oriented\end{tabular} & mutability &
\begin{tabular}[c]{@{}l@{}}system\\ based\end{tabular} & XML & XACML &\_ & \_\\ 

\rowcolor[HTML]{EFEFEF}
\citet{9161997,8717790}
& A, O & \begin{tabular}[c]{@{}l@{}}environmental\\ system-oriented\end{tabular} & mutability &
\begin{tabular}[c]{@{}l@{}}system\\ based\end{tabular} & XML & XACML &\_ & \_\\

\citet{8029555}
& A, O & \begin{tabular}[c]{@{}l@{}}environmental\\ system-oriented\end{tabular} & mutability &
\begin{tabular}[c]{@{}l@{}}system\\ based\end{tabular} & XML & XACML &\_ & \_\\ 

\rowcolor[HTML]{EFEFEF}
\citet{Marra2019}
& A, O & \begin{tabular}[c]{@{}l@{}}environmental\\ system-oriented\end{tabular} & mutability &
\begin{tabular}[c]{@{}l@{}}system\\ based\end{tabular} & XML & XACML &\_ & \_\\

\citet{10.1007/978-3-319-72389-1_43}
& A, O & \begin{tabular}[c]{@{}l@{}}environmental\\ system-oriented\end{tabular} & mutability &
\begin{tabular}[c]{@{}l@{}}system\\ based\end{tabular} & XML & XACML &\_ & \_\\ 

\rowcolor[HTML]{EFEFEF}
\citet{10.1145/2851613.2851797}
& A, O & \begin{tabular}[c]{@{}l@{}}environmental\\ system-oriented\end{tabular} & mutability & 
\begin{tabular}[c]{@{}l@{}}system\\ based\end{tabular} & XML & XACML &\_ & \_ \\

\citet{Martinelli2019ObligationMI} &
  A, O &
  \begin{tabular}[c]{@{}l@{}}environmental\\ system-oriented\end{tabular} &
  mutability &
  \begin{tabular}[c]{@{}l@{}}system\\ based\end{tabular} &
  XML &
  XACML &
  \_ & \_\\ 

\rowcolor[HTML]{EFEFEF}
\citet{su12093885,MunozArcentales2019AnAF} &
  P, Pr, O &
  \begin{tabular}[c]{@{}l@{}}temporal\\ spatial\\ amount-based\end{tabular} &
  \_ &
  \_ &
  XML &
  ODRL &
  \_& \_ \\ 

\citet{6045968}
  &
  E, C, Ac &
  \begin{tabular}[c]{@{}l@{}}cardinal\\ temporal\\ spatial\end{tabular} &
  mutability &
  \_ &
  XML &
  \_ &
  \_ & OSL\\ 

\rowcolor[HTML]{EFEFEF}
\citet{5319024} & S, T, Ac
   & \begin{tabular}[c]{@{}l@{}} temporal\\ spatial\end{tabular} &mutability
   & \begin{tabular}[c]{@{}l@{}}model\\ based\end{tabular} 
   & \_
   & \_
   &\_& \_
   \\ 
   
\citet{Schtte2018LUCONDF} &
  Al, D, O &
  \_ &
   &
  \_ &
  Java &
  DSL &
  \_ & first order logic \\ 
\rowcolor[HTML]{EFEFEF}
\citet{10.1016/j.jnca.2011.03.019} &
  A, O &
  \begin{tabular}[c]{@{}l@{}}environmental\\ system-oriented\end{tabular} &
  mutability &
  \_ &
  LALR &
  UCON &
  \_ & \_\\ 
  
\citet{Silva2010}
  &
  E, C, Ac &
  \begin{tabular}[c]{@{}l@{}}cardinal\\ temporal\\ spatial\end{tabular} &
  \_ &
  \_ &
  XML &
  \_ &
  \_ & OSL\\ 
  
\rowcolor[HTML]{EFEFEF}  
\citet{Wchner2013CompliancePreservingCS}
  &
  E, C, Ac &
  \begin{tabular}[c]{@{}l@{}}cardinal\\ temporal\\ spatial\end{tabular} &
  \_ &
  \_ &
  XML &
  \_ &
  \_ & OSL\\ 

  \citet{6405363}
  &
  E, C, Ac &
  \begin{tabular}[c]{@{}l@{}}cardinal\\ temporal\\ spatial\end{tabular} &
  \_ &
  \_ &
  XML &
  \_ &
  \_ & OSL\\ 

\rowcolor[HTML]{EFEFEF}
\citet{Xu2007} &
  E, P, Ac &
  \_ &
  mutability &
  \_ &
  XML &
  UCON &
  \_ & \_\\ 

\citet{Zhang2008TowardAU} &
  A, O &
  \begin{tabular}[c]{@{}l@{}}environmental\\ system-oriented\end{tabular} &
  mutability &
  \begin{tabular}[c]{@{}l@{}}model\\ based\end{tabular} &
  XML &
  UCON &
  PEI & \_\\ 
  
\end{tabular}
\label{tab:specification}
\end{table}

\FloatBarrier
\footnotetext{P= permission; Pr= prohibition; O= obligation; D= dispensation ; A= authorizations; E= event; C= conditions; P= predicate; Ac= actions, Al= allow; D= drop; S= Subject; T= Target; For attributes, if 'mutability' then the policy supports mutable attributes, if blank then the policy doesn't support mutable attributes, else the policy does not support attributes at all; For the rest of the columns, if '-', then not supported}

\paragraph{Operators/ Rules.}
The policy languages presented in \cite {10.1007/978-3-642-16161-2_19,Zhang2008TowardAU,10.1016/j.jnca.2011.03.019,Xu2007} support the core UCON \cite{10.1145/984334.984339} model components, namely,  authorizations and obligations. 
UCON\textsubscript{KI}, a policy language proposed by \citet{Xu2007},
which only supports authorizations, is an event-based UCON model that uses the Event-Predicate-Action (EPA) language that originated from Event-Condition-Action (ECA) \cite{Alferes2006} rules. An event is an activity carried out by a subject; the action part consists of preventive and detective mechanisms; and the predicate part defines UCON authorizations.

The frameworks presented in 
\cite{Lazouski2012APF,6470943,CARNIANI201637,10.1007/978-3-319-68063-7_8,9161997,8717790,8456147,8029555,Marra2019,10.1007/978-3-319-72389-1_43, Martinelli2019ObligationMI,10.1145/2851613.2851797}
use the U-XACML policy language and model, which was originally proposed by \citet{10.1007/978-1-4419-6794-7_11}. U-XACML is an extension of XACML that introduces attribute updates and continuous policy evaluation. It mainly supports UCON authroizations and XACML obligations. In the UCON model, obligations are actions that have to be performed by subjects. Whereas, in XACML obligations have different semantics and are considered as duties that are performed by the enforcement mechanism in order to enforce access decisions \cite {Lazouski2010UsageCI}. Later \citet{Martinelli2019ObligationMI} extended U-XACML in order to provide support for UCON obligations.
A similar policy language to U-XACML is OB-XACML, which is proposed by \citet{kateb2014}. However, OB-XACML is able to formalize post-obligations and contextual information via the OB-XACML model. 

\citet{CAO2020998} propose a \emph{data usage control model}, entitled DUPO, that can be used to cater for diverse usage control policies. The policy language is based on defeasible logic and enriched with deontic operators. Another policy language that uses deontic operators to express usage control policies is the \emph{open digital rights language} (ODRL) \cite{ODRL2018}, which is  used by both \citet{MunozArcentales2019AnAF,su12093885} and \citet{9314823}. ODRL is a 
World Wide Web Consortium (W3C)\footnote{W3C, https://www.w3.org/} standard that provides an information model, vocabulary, and encoding mechanisms that can be used to represent statements about content and services usage\footnote{ODRL, https://www.w3.org/community/odrl/}.

The policy language adopted by \cite{6045968,6405363,Wchner2013CompliancePreservingCS,Silva2010,6258293, Jung2014,Baldini2013,Neisse2015SecKitAM} uses ECA rules to express policies. 
Conditions, which are mainly used to impose temporal and cardinal constraints on data usage, are expressed using the \emph{obligation specification language} (OSL) \cite{10.1007/978-3-540-74835-9_35}. The formal model of OSL is logic based and specified in the Z language \cite{DBLP:conf/ds/Abrial74}. The action part consists of preventive and detective mechanisms. Unlike the original UCON model, OSL is able to formalize, alongside pre-obligations and ongoing-obligations, post-obligations \cite{Lazouski2010UsageCI}.
%

The policy language proposed by \citet{Schtte2018LUCONDF}, which is represented in a \emph{domain specific language} (DSL), named LUCON DSL, is specified in Java and compiled into first order logic. The DSL grammar is used to express the rules that determine whether the sending of a message is allowed or prohibited.
This grammar also allows a decision to be linked to an obligation, in particular a pre-obligation.
%
The policy language underpinning the xDUCON framework proposed by \citet{5319024} is represented in xDSpace, an implementation of the \emph{shared data space} \cite{Gelernter1985GenerativeCI} programming system. xDSpace allows rules to be expressed as tuples that include subjects, targets (or objects), and actions to be performed on the object by the subject.

\paragraph{Conditions.}
UCON-based policy languages \cite{10.1007/978-3-642-16161-2_19,Zhang2008TowardAU,10.1016/j.jnca.2011.03.019,Xu2007,kateb2014, Lazouski2012APF,6470943,CARNIANI201637,10.1007/978-3-319-68063-7_8, 9161997,8717790,8456147,8029555,Marra2019,10.1007/978-3-319-72389-1_43,Martinelli2019ObligationMI,10.1145/2851613.2851797}
allow different environmental and system-based conditions to be expressed. However, the  policy language proposed by \citet{10.1016/j.jnca.2011.03.019} only caters for four predefined conditions, namely, the current time, the amount of CPU in use, and the amount of free memory and disk space.
ODRL \cite{9314823,su12093885,MunozArcentales2019AnAF} allows various conditions to be expressed, such as, temporal, spatial, amount-based, purpose, and event-defined.
The DUPO policy language presented in \cite{CAO2020998} also allows event-defined conditions to be expressed. 
DUPO allows five pre-defined conditions to be expressed, namely, actor, spatial, temporal, purpose and monetization.
The OSL-based policy languages presented in \cite{6045968,6405363,Wchner2013CompliancePreservingCS,Silva2010,6258293, Jung2014,Baldini2013,Neisse2015SecKitAM} are mainly used to express temporal and cardinal conditions on the data using OSL. Moreover, the policy language used in the  framework proposed by \citet{6258293}
allows additional attributes defined as XML spatial conditions to be captured. Whereas, the  policy language proposed by \citet{Jung2014} makes use of the condition part to refer to additional attributes such as contextual information relating to the policy. 
The xDUCON policy language \cite{5319024} allows for the expression of two predefined conditions that represent the contextual information captured by the framework, namely, temporal and spatial conditions.

\paragraph{Attributes.} 
Most of the UCON-based policy languages presented in \cite{10.1007/978-3-642-16161-2_19,Zhang2008TowardAU,10.1016/j.jnca.2011.03.019,Xu2007,Lazouski2012APF,6470943,CARNIANI201637,10.1007/978-3-319-68063-7_8,9161997,8717790,8456147,8029555,Marra2019,10.1007/978-3-319-72389-1_43,Martinelli2019ObligationMI,10.1145/2851613.2851797}
support attribute mutability by allowing attribute updates. The authors of the framework presented in \cite{10.1007/978-3-642-16161-2_19} detailed the policy attributes responsible for controlling ongoing obligations and attributes using the XML policy attribute \emph{obligationTime} and an \emph{update policy}, respectively, which indicates the attribute to be updated and the update time. In addition, the U-XACML policy language presented in \cite{Lazouski2012APF,6470943,CARNIANI201637, 10.1007/978-3-319-68063-7_8,9161997,8717790,8456147,8029555,Marra2019, 10.1007/978-3-319-72389-1_43,Martinelli2019ObligationMI,10.1145/2851613.2851797} supports attribute mutability by introducing \emph{AttrUpdate} in the policy specification. The framework proposed by \citet{6045968} allows attributes to be updated by exploiting the \emph{attributeMatch} element in XML, which is bound to temporal operators. The attributes supported by the various policy languages depend significantly on the scope of the usage control solution.

\paragraph{Context.}
Several authors \cite{Jung2014,CAO2020998,10.1007/978-3-642-16161-2_19, Zhang2008TowardAU,5319024,kateb2014,Baldini2013,Neisse2015SecKitAM} introduced contextual information in their policy specification. 
For instance, the policy languages adopted by \citet{Zhang2008TowardAU} and \citet{Jung2014} use conditions to support context-based authorizations. Another framework, the one proposed by \citet{10.1007/978-3-642-16161-2_19} presents the ConUCON policy language based on the ConUCON policy model, which is an extension of UCON with a new component that caters for contextual information. The new context component allows ongoing environmental (e.g., spatial and temporal) and system (e.g., CPU and battery) information to be captured. 
The remaining frameworks presented in \cite{Lazouski2012APF,6470943,CARNIANI201637,10.1007/978-3-319-68063-7_8, 9161997,8717790,8456147,8029555,Marra2019,10.1007/978-3-319-72389-1_43,Martinelli2019ObligationMI,10.1145/2851613.2851797} are context based-systems, as they directly make use of the contextual information collected via their system components.

\subsubsection{Flexibility \& Extensibility.} Some policies owe their flexibility and extensibility to the conceptual model that expresses the policy in an abstract way and/or to the language used to encode policies.
\paragraph{Representation Format.}
Most of the policy languages are represented in XML.
The remaining ones are represented either in DSL specified in Java \cite{Schtte2018LUCONDF} or the \emph{look-ahead-left-to-right} (LALR) grammar \cite{10.1016/j.jnca.2011.03.019}.
\citet{Zhang2008TowardAU} state that the XML language is extensible enough to meet the expressiveness and flexibility of the UCON model, but also its extensions. Besides, \citet{6258293} agree that XML allows new policy rules to be added in order to express and extend different models, which is confirmed by \citet{10.1007/978-1-4419-6794-7_11} who successfully incorporated UCON components in the XACML policy language. 
For the remaining representation languages, \citet{10.1016/j.jnca.2011.03.019} and \citet{Schtte2018LUCONDF} assert that their proposed languages, which use the LALR grammar and Java, respectively, are sufficiently extensible and flexible to meet new conditions and usage rules.

\paragraph{Conceptual Model.}
The UCON model has demonstrated great flexibility as it has been used in different application contexts, such as industry 4.0, operating systems, and mobile computing. 
For example, \citet{10.1007/978-3-642-16161-2_19} claim that the ConUCON policy model can be implemented not only for Android but also for other mobile platforms due to the policy flexibility and extensibility. Moreover, \citet{Kattinproceedings2008} incorporated post-obligations in the context of an Industry 4.0 application and \citet{10.1007/978-3-642-16161-2_19} extended the original UCON model with context components in order to develop context-aware ubiquitous systems.
As for ODRL, according to \citet{MunozArcentales2019AnAF} and \citet{9314823}, the policy language presents a flexible policy model, which allows various usage scenarios to be expressed, but also a fully extensible model, which provides mechanisms to extend and/or deprecate the original model. 
Regarding XACML, although the model was originally used for access control specification, it has shown a high level of flexibility and extensibility to support UCON components, as indicated by \citet{10.1007/978-1-4419-6794-7_11}. Besides, the framework proposed by \citet{Lazouski2012APF} and the framework proposed by \citet{6470943} and \citet{CARNIANI201637} and its various extensions presented in
\cite{10.1007/978-3-319-68063-7_8,9161997,8717790,8456147,8029555,Marra2019,10.1007/978-3-319-72389-1_43,Martinelli2019ObligationMI,10.1145/2851613.2851797}
have underlined the degree of extensibility of XACML as it has undergone various extensions over time in order to cater for different types of attributes (e.g., attributes that account for different sensors in a IoT scenario or describe the features of subjects, resources, and environment change) and conditions (e.g., location, time, occurrence, and event-based). 
\subsubsection{Unambiguous.}
\citet{Zhang2008TowardAU} describe the highest level of policy (i.e., the high level objectives of normative policies) as being informal and fuzzy, which makes it difficult to enforce them effectively. In their work, the authors developed their framework using the layered \emph{policy-enforcement-implementation} (PEI) \cite{10.1145/1128817.1128820} methodology. PEI seeks to bridge the gap between informal or high-level policies and the actual enforcement mechanism, thereby undermining the ambiguity of informal usage control policies. PEI is composed of five layers: security and system goals, policy models, enforcement models, implementation models, and actual implementations.
The first layer is necessarily informal, while the second layer aims to take high-level informal goals and provide concrete details using formal or quasi-formal notation. For the authors, the UCON  model presents the formal layer of their policy language, while  enforcement and implementation models are associated with the actual code that implements the solution. 
The designers of the remaining frameworks do not mention any methodology or formal ways of dealing with the unambiguity requirement.

\subsubsection{Formal Semantics.}
Logic-based approaches that formalize the proposed policy languages were employed by  \cite{Jung2014,6045968,6405363,Wchner2013CompliancePreservingCS,Silva2010,6258293,Baldini2013,Neisse2015SecKitAM}, which use OSL, and the ones presented in \cite{CAO2020998,Schtte2018LUCONDF}, which use defeasible logic and first order logic, respectively. While the remaining policy languages  such as U-XACML and ODRL do not provide any formal foundations, some works such as \cite{Martinelli2016EnforcementOU,Polleres2015} propose ways to formalize these languages.

\begin{table}[t!]
\tiny
\centering
\caption{Usage Control Framework Components}
\begin{tabular}{p {4 cm}p {3cm}p {5 cm}}
\rowcolor[HTML]{C0C0C0} 
 \textbf{Framework} &
  \textbf{Type} &
  \textbf{Components} 
  \\ 
\citet{10.1007/978-3-642-16161-2_19} &
  XACML &
  \begin{tabular}[c]{@{}l@{}}PEP, PDP, PIP,  PAP,  \\ evaluation engines\end{tabular}  \\ 

\rowcolor[HTML]{EFEFEF} 
\citet{Baldini2013,Neisse2015SecKitAM} &
 XACML &
  \begin{tabular}[c]{@{}l@{}}PEP, PDP, \\ context manager,  \\ role manager,  \\ policy management,\\ graphical user interface, \\ policy repository,\\ policy server\end{tabular}  \\ 

\citet{CAO2020998} &
  custom &
  \begin{tabular}[c]{@{}l@{}}identification,    \\ policy management,    \\ policy composition,   \\ data usage transparency,  \\ data usage traceability, \\jDUPO \end{tabular}  \\ 

\rowcolor[HTML]{EFEFEF} 
\citet{CARNIANI201637,6470943} &
 XACML &
\begin{tabular}[c]{@{}l@{}}PEP, PIP, PDP, PAP,  \\ context handler, attribute manager,\\ session manager\end{tabular}  \\ 

\citet{9314823} &
 XACML &
 \begin{tabular}[c]{@{}l@{}}policy management, PEP, \\graphical user interface,  \\ infrastructure services\end{tabular} \\ 

\rowcolor[HTML]{EFEFEF} 
\citet{8456147} &
  XACML &
  \begin{tabular}[c]{@{}l@{}}PEP, PIP, PDP, PAP,  \\ context handler, usage monitor\end{tabular}  \\ 

\citet{6258293} &
  XACML &
  PEP, PDP\\ 

\rowcolor[HTML]{EFEFEF} 
\citet{10.1007/978-3-319-68063-7_8} &
  XACML &
  \begin{tabular}[c]{@{}l@{}}PEP, PIP, PDP, PAP,\\ context handler, attribute manager,\\ session manager\end{tabular} 
  \\ 
  
\citet{Jung2014} &
  XACML &
  \begin{tabular}[c]{@{}l@{}}PXP, PEP, PRP, \\ PDP, PIP, PMP, PAP\end{tabular}  \\ 
  
\rowcolor[HTML]{EFEFEF} 
\citet{kateb2014} &
  XACML &
  \begin{tabular}[c]{@{}l@{}}PIP, PDP, \\ obligation manager\end{tabular} \\ 

\citet{Lazouski2012APF} &
  XACML &
  \begin{tabular}[c]{@{}l@{}}PEP, PIP, PDP, PAP,  \\ context handler, usage monitor\end{tabular} 
  \\ 

\rowcolor[HTML]{EFEFEF} 
\citet{9161997,8717790} &
  XACML &
  \begin{tabular}[c]{@{}l@{}}PEP, PIP, PDP, PAP,  \\ context handler, usage monitor\end{tabular}
   \\ 

\citet{8029555} &
 XACML &
 \begin{tabular}[c]{@{}l@{}}PEP, PIP, PDP, PAP,  \\ context handler, usage monitor\end{tabular} \\ 

\rowcolor[HTML]{EFEFEF}  
\citet{Marra2019} &
XACML &
  \begin{tabular}[c]{@{}l@{}}PEP, PIP, PDP, PAP,\\ context handler, usage monitor\end{tabular} \\ 

\citet{10.1007/978-3-319-72389-1_43} &
 XACML &
\begin{tabular}[c]{@{}l@{}}PEP, PIP, PDP, PAP,\\ context handler, usage monitor\end{tabular} \\ 

\rowcolor[HTML]{EFEFEF} 
\citet{Martinelli2019ObligationMI} &
  XACML &
  \begin{tabular}[c]{@{}l@{}}PEP, attribute manager, PIP, PDP,   \\ PAP, session manager, context handler,\\ OEP, OOP, POP, ODP\end{tabular} 
   \\ 
   
\citet{10.1145/2851613.2851797} &
 XACML &
  \begin{tabular}[c]{@{}l@{}}PEP, PIP, PDP, PAP,\\ context handler, usage monitor\end{tabular} 
  \\ 

\rowcolor[HTML]{EFEFEF}  
\citet{su12093885,MunozArcentales2019AnAF} &
  XACML &
 PDP, PXP, PTP  \\ 
 
\citet{6045968} &
  XACML &
  \begin{tabular}[c]{@{}l@{}}policy manager, \\ policy repository, PDP,   \\ attribute resolver,   \\ action resolver, PEP\end{tabular}\\ 

\rowcolor[HTML]{EFEFEF} 
\citet{5319024}  &
  XACML &
  PEP, PDP, xDSpace
 \\ 
 
\citet{Schtte2018LUCONDF} &
  custom &
  \begin{tabular}[c]{@{}l@{}}interceptor,\\ Prolog engine\end{tabular} \\ 
  
\rowcolor[HTML]{EFEFEF} 
\citet{10.1016/j.jnca.2011.03.019} &
 custom &
  \begin{tabular}[c]{@{}l@{}}reference monitor, \\ usage mediator, \\ rule parser\end{tabular} 
   \\

\citet{Silva2010} &
  custom &
\begin{tabular}[c]{@{}l@{}}policy editor,   \\ notification manager,  \\ policy monitor,  \\ control monitor,  \\ event signaler\end{tabular} 
 \\ 

\rowcolor[HTML]{EFEFEF} 
\citet{Wchner2013CompliancePreservingCS} &
  XACML &
  PEP, PIP, PDP \\ 
  
\citet{6405363} &
  XACML &
  PEP, PIP, PDP \\ 

\rowcolor[HTML]{EFEFEF} 
\citet{Xu2007} &
  XACML &
  enforcer, attribute repository, PDP 
   \\

\citet{Zhang2008TowardAU} &
 XACML &
  PDP, PEP, usage monitor, attribute repository \\

\end{tabular}%
\label{tab:framework-components}
\end{table}

\subsection{Enforcement}
\begin{table}[t!]
\tiny
\centering
\caption{Usage Control Enforcement Mechanisms}
\begin{tabular}{p {2.2cm}p {2cm}p {1.5 cm}p{2.8 cm}p{1.5 cm}p{1.4cm}}

\rowcolor[HTML]{C0C0C0} 
 \textbf{Framework} &
  \textbf{Preventive} &
  \textbf{Detective} &
  \textbf{Continuity of Enforcement} &
  \textbf{Conflict Detection \& Resolution} 
  &
  \textbf{Administration} 
  \\ 

\citet{10.1007/978-3-642-16161-2_19} &
  permission, inhibition,  revoke, delay, update &
  \_ &
  monitor attribute and context updates, monitor obligation fulfillment & \_
   & PAP interface
  \\  

  \rowcolor[HTML]{EFEFEF} 
\citet{Baldini2013,Neisse2015SecKitAM} &permission, inhibition, revoke, delay, modification &\_ 
&monitor condition updates and obligations fulfillment
&combining algorithms 
  & graphical user interface 
  \\ 

 \citet{6470943,CARNIANI201637}
  &
  permission, inhibition,  revoke, update  &
  \_ &
  monitor attribute, context, and condition updates &
  \_
  & PAP interface
  \\ 

  \rowcolor[HTML]{EFEFEF}  
\citet{CAO2020998} &
  permission, inhibition,  revoke, delay &
  \_ &
  monitor condition and context updates, monitor obligation fulfillment &
  logic-based
  & jDUPO
  \\ 
 
\citet{8456147}
  &
  permission, inhibition,  revoke, update  &
  \_ &
  monitor attribute, context, and condition updates &
  \_ 
  & PAP interface
  \\ 

  \rowcolor[HTML]{EFEFEF} 
\citet{9314823} &
  permission, inhibition, execution &
  execution actions &
  \_ &
  & graphical user interface
   \\

 \citet{6258293} &
 permission, inhibition, revoke, delay, modification, execution &
  execution actions &
  monitor condition updates and obligations fulfillment &
  \_
  &Android interface
  \\ 
 
 \rowcolor[HTML]{EFEFEF}  
 \citet{10.1007/978-3-319-68063-7_8}
  & 
  permission, inhibition,  revoke, update  &
  \_ &
  monitor attribute, context, and condition updates &
  \_ 
  & PAP interface
  \\   
  
\citet{Jung2014} &
  permission, inhibition, modification, execution, revoke &
  execution actions &
  monitor condition updates and obligations fulfillment &
  \_ 
  & PAP interface
  \\ 

  \rowcolor[HTML]{EFEFEF}   
 \citet{kateb2014} & permission, inhibition, revoke, delay & \_ & monitor obligation fulfillment &\_ 
  &\_
  \\ 
  
 \citet{9161997,8717790}
  &
  permission, inhibition,  revoke, update  &
  \_ &
  monitor attribute, context, and condition updates &
  \_ 
  & PAP interface
  \\ 

  \rowcolor[HTML]{EFEFEF}  
 \citet{Lazouski2012APF}
  &
  permission, inhibition,  revoke, update  &
  \_ &
  monitor attribute, context, and condition updates &
  \_ 
  &PAP interface
  \\ 

\citet{8029555}
  &
  permission, inhibition,  revoke, update  &
  \_ &
  monitor attribute, context, and condition updates &
  \_ 
  & PAP interface
  \\ 
 
   \rowcolor[HTML]{EFEFEF} 
\citet{Marra2019}
  &
  permission, inhibition,  revoke, update  &
  \_ &
  monitor attribute, context, and condition updates &
  \_ 
  & PAP interface
  \\ 

\citet{10.1007/978-3-319-72389-1_43}
  &
  permission, inhibition,  revoke, update  &
  \_ &
  monitor attribute, context, and condition updates &
  \_ 
  & PAP interface
  \\ 

\rowcolor[HTML]{EFEFEF} 
\citet{10.1145/2851613.2851797} &
  permission, inhibition, suspend, resume, update &
  \_ &
  monitor attribute, context, and condition updates &
  \_ 
  & PAP interface
  \\ 
  
\citet{Martinelli2019ObligationMI} &
  permission,  inhibition, revoke, delay, update &
  \_ &
  monitor attribute, context, condition and updates, monitor obligation fulfillment &
  \_ 
  &PAP interface
  \\ 
  
\rowcolor[HTML]{EFEFEF} 
\citet{su12093885,MunozArcentales2019AnAF}  &
  permission, inhibition, revoke, delay, modification &
  \_ &
  monitor condition and obligation fulfillment &
  \_ 
  & PAP interface
  \\ 

\citet{6045968} &
 permission, inhibition, revoke, delay, modification, execution, update &
  execution actions &
  monitor condition and attributes updates, monitor obligations fulfillment &
  \_ 
  &\_
  \\ 

\rowcolor[HTML]{EFEFEF} 
\citet{5319024} & permission, inhibition, revoke, delay, update &\_ &monitor condition and context updates, monitor obligation fulfillment &\_ 
  &\_
  \\ 
  
\citet{Schtte2018LUCONDF} &
  permission, inhibition, modification &
  auditing &
  monitor data flow &
  logic-based 
  & Eclipse integrated development environment
  \\ 

\rowcolor[HTML]{EFEFEF} 
\citet{10.1016/j.jnca.2011.03.019} &
  permission, inhibition, revoke, update &
  \_ &
  monitor attribute updates &
  \_ 
  &\_
  \\ 
  
\citet{Silva2010} &
 permission, inhibition, revoke, delay, modification, execution &
  execution actions &
  monitor condition updates and obligations fulfillment &
  \_ 
  & policy editor
  \\ 
 
\rowcolor[HTML]{EFEFEF} 
 \citet{6405363} &
 permission, inhibition, revoke, delay, modification, execution &
  execution actions &
  monitor condition updates and obligations fulfillment &
  \_ 
  & \_
  \\ 

 \citet{Wchner2013CompliancePreservingCS} &
 permission, inhibition, revoke, delay, modification, execution &
  execution actions &
 monitor condition updates and obligations fulfillment &
  \_ 
  & \_
  \\ 
  
 \rowcolor[HTML]{EFEFEF} 
\citet{Xu2007} &
  permission, inhibition, revoke, update &
  \_ &
  monitor attribute updates &
  \_ 
  &LINUX command line
  \\ 
  
\citet{Zhang2008TowardAU} &
  permission, inhibition, revoke, update, delay &
  \_ &
  monitor attribute updates and obligation fulfillment &
  \_ 
  & \_
  \\ 
  \rowcolor[HTML]{EFEFEF}

\end{tabular}%

\label{tab:enforcement}
\end{table}
In Table \ref{tab:framework-components}, we outline the various components employed in each framework, while in
Table \ref{tab:enforcement}, we present a comparative overview of the predominant usage control frameworks found in the literature. 
Most of  the frameworks depicted in Table \ref{tab:framework-components} are XACML \cite{XACML2013} reference architectures, and thus include some or all of the following components: a \emph{policy decision point} (PDP) \cite{10.1007/978-3-642-16161-2_19,Baldini2013,Neisse2015SecKitAM,CARNIANI201637,6470943,8456147,6258293,10.1007/978-3-319-68063-7_8,Jung2014,kateb2014,Lazouski2012APF,9161997,8717790,8029555,Marra2019,10.1007/978-3-319-72389-1_43,Martinelli2019ObligationMI,10.1145/2851613.2851797,su12093885,MunozArcentales2019AnAF,6045968,5319024,Wchner2013CompliancePreservingCS,6405363,Xu2007,Zhang2008TowardAU}, a \emph {policy enforcement point} (PEP) \cite{10.1007/978-3-642-16161-2_19,Baldini2013,Neisse2015SecKitAM,CARNIANI201637,6470943,9314823,8456147,6258293,10.1007/978-3-319-68063-7_8,Jung2014,Lazouski2012APF,9161997,8717790,8029555,Marra2019,10.1007/978-3-319-72389-1_43,Martinelli2019ObligationMI,10.1145/2851613.2851797,6045968,5319024,Wchner2013CompliancePreservingCS,6405363,Zhang2008TowardAU}, a \emph{policy information point} (PIP) \cite{10.1007/978-3-642-16161-2_19,CARNIANI201637,6470943,8456147,10.1007/978-3-319-68063-7_8,Jung2014,kateb2014,Lazouski2012APF,9161997,8717790,8029555,Marra2019,10.1007/978-3-319-72389-1_43,Martinelli2019ObligationMI,10.1145/2851613.2851797,Wchner2013CompliancePreservingCS,6405363}, a \emph{policy execution point} (PXP) \cite{Jung2014,su12093885,MunozArcentales2019AnAF}, a \emph{policy administration point} (PAP) \cite{10.1007/978-3-642-16161-2_19,CARNIANI201637,6470943,8456147,10.1007/978-3-319-68063-7_8,Jung2014,Lazouski2012APF,9161997,8717790,8029555,Marra2019,10.1007/978-3-319-72389-1_43,Martinelli2019ObligationMI,10.1145/2851613.2851797}, a \emph{policy retrieval point} (PRP) \cite{Jung2014},  a \emph{policy management point} (PMP) \cite{Jung2014}, and/or a \emph{policy translation point} (PTP) \cite{su12093885,MunozArcentales2019AnAF}.
Although the vast majority of frameworks
introduce usage control extensions for XACML, some frameworks \cite{CAO2020998,Schtte2018LUCONDF,10.1016/j.jnca.2011.03.019,Silva2010} also introduce novel enforcement mechanisms. 

Various extensions of XACML include components for the continuous evaluation and the enforcement of usage control policies. Existing proposals include a variety of new components that are responsible for managing and evaluating policy attributes, namely an \emph{attribute manager} \cite{CARNIANI201637,6470943,10.1007/978-3-319-68063-7_8,Martinelli2019ObligationMI}, a \emph{session manager} \cite{Martinelli2019ObligationMI,CARNIANI201637,6470943,10.1007/978-3-319-68063-7_8}, a \emph{usage monitor} \cite{8456147,Lazouski2012APF,9161997,8717790,8029555,Marra2019,10.1007/978-3-319-72389-1_43,10.1145/2851613.2851797,Zhang2008TowardAU}, an \emph{attribute repository} \cite{Xu2007,Zhang2008TowardAU}, a \emph{role manager} \cite{Baldini2013,Neisse2015SecKitAM}, and  an \emph{evaluation engine} \cite{10.1007/978-3-642-16161-2_19}. Additionally, there are various proposals for managing and evaluating policy obligations via an \emph{obligation manager} \cite{kateb2014}, an \emph{evaluation engine} \cite{10.1007/978-3-642-16161-2_19}, a \emph{session manager} \cite{Martinelli2019ObligationMI,CARNIANI201637,6470943,10.1007/978-3-319-68063-7_8}, an \emph{action resolver} \cite{6045968}, an \emph{obligation enforcement point} (OEP) \cite{Martinelli2019ObligationMI}, an \emph{obligation observation point} (OOP) \cite{Martinelli2019ObligationMI}, and a \emph{policy obligation point} (POP) \cite{Martinelli2019ObligationMI}.
When it comes to managing contextual information, extensions include a \emph{context handler} \cite{CARNIANI201637,6470943,8456147,10.1007/978-3-319-68063-7_8,Lazouski2012APF,9161997,8717790,8029555,Marra2019,10.1007/978-3-319-72389-1_43,Martinelli2019ObligationMI,10.1145/2851613.2851797}, a \emph{context manager} \cite{Baldini2013,Neisse2015SecKitAM}, a \emph{session manager} \cite{Martinelli2019ObligationMI,CARNIANI201637,6470943,10.1007/978-3-319-68063-7_8}, and an \emph{evaluation engine} \cite{10.1007/978-3-642-16161-2_19}.
Whereas, \emph{policy management} \cite{Baldini2013,Neisse2015SecKitAM,CAO2020998,9314823}, \emph{policy manager} \cite{6045968}, and \emph{policy repository} \cite{6045968,Baldini2013,Neisse2015SecKitAM} components have been proposed in order to support the management of usage control policies. 
Finally, there are a number of suggestions for managing the communication between various components, including an \emph{infrastructure service} \cite{9314823} that can be used as a means to develop trust and a \emph{shared data space} (xDSpace) \cite{5319024} that can be used as a tool for coordinating the execution of distributed applications.

When it comes to the novel frameworks, various architectural designs are motivated by different aims and requirements.
For instance, the \emph{trustworthy data sharing platform} \cite{CAO2020998} includes five predominant features:   \emph{identification},  \emph{policy management} (visualization tool), \emph{policy composition}, \emph{data usage transparency}, and \emph{data usage traceability}.
The framework proposed by 
\citet{Schtte2018LUCONDF} uses a \emph{Prolog engine} as a PEP and an \emph{interceptor for events} as a PDP.
The  enforcement architecture proposed  by \citet{10.1016/j.jnca.2011.03.019} is based on three main components: a \emph{reference monitor} that acts as a PDP, a \emph{usage mediator} that acts as a PEP, and a \emph {LALR rule parser} that translates the rules expressed by the LALR grammar into an internal representation, which is used by the \emph{reference monitor}.
The framework proposed by \citet{Silva2010} includes a \emph {control monitor} and a \emph{policy monitor} that together act as a PDP, an \emph{event signaler} that acts as a PEP, and a \emph{notification manager} that sends notifications to the \emph{control monitor} in the case of violations.
\subsubsection{Preventive.}
The preventive mechanisms of a variety of frameworks are provided by the enforcement points.
With the exception of the frameworks described in \cite{10.1145/2851613.2851797}, \cite{9314823} and \cite{Schtte2018LUCONDF}, each of the frameworks can enforce: \emph{permissions}, \emph{prohibitions}, and \emph{revocations}.
These classes of enforcement are responsible for enforcing usage decisions by allowing, denying access to or usage of a resource, and revoking access in the case of policy violations. 
While, the \emph{suspend} and \emph{resume} framework proposed by \citet{10.1145/2851613.2851797} allows access to be resumed after suspension by the system as a result of an ongoing-evaluation. The frameworks proposed in \cite{9314823} and \cite{Schtte2018LUCONDF} do not allow access to data to be revoked. 
Another class of enforcement is \emph{modification}. The frameworks presented in \cite{Jung2014,Schtte2018LUCONDF,su12093885,MunozArcentales2019AnAF,6045968,6405363,Wchner2013CompliancePreservingCS,Silva2010,6258293,Baldini2013,Neisse2015SecKitAM} modify certain data values after access is granted in order to allow the user to use the data while ensuring privacy protection and policy compliance. 
Only a few frameworks \cite{6045968,6405363,Wchner2013CompliancePreservingCS,Silva2010,6258293,Jung2014,9314823} cater for the \emph{execution} of actions or non-usage actions that trigger required actions to be performed (e.g., sending a notification, triggering a payment, or writing in logs).
\citet{9314823} state that their framework enforces three main types of actions, namely, anonymization of the data before use by the data consumer; making sure that the data never leave the providers domain; and deleting traces of data usage (e.g., search results) in the data consumers domain-memory.
The frameworks proposed in \cite{CAO2020998,10.1007/978-3-642-16161-2_19, Zhang2008TowardAU,Martinelli2019ObligationMI, su12093885,MunozArcentales2019AnAF, kateb2014,5319024,Baldini2013,Neisse2015SecKitAM} provide mechanisms that allow for the enforcement of obligation fulfillment, which is usually referred to as a \emph{delay} class of enforcement, as the framework delays the access to or the usage of a resource until users  perform certain obligations, which in turn trigger the re-evaluation of the relevant policies.
The framework proposed by \citet{Zhang2008TowardAU} only enforces pre-obligations, whereas the rest can enforce pre-obligations, on-obligations, and post-obligations.
The frameworks presented in \cite{Lazouski2012APF, 6470943,CARNIANI201637,10.1007/978-3-319-68063-7_8,9161997,8717790,8456147,8029555,Marra2019,10.1007/978-3-319-72389-1_43, 10.1007/978-3-642-16161-2_19,Schtte2018LUCONDF,Xu2007,10.1145/2851613.2851797} only consider XACML-type obligations that are handled internally. While, the framework presented in \cite{10.1016/j.jnca.2011.03.019} considers obligations as external to the decision system and thus said obligations are managed by external routines.
The various frameworks that are based on the UCON model
\cite{Lazouski2012APF,6470943,CARNIANI201637,10.1007/978-3-319-68063-7_8, 9161997,8717790,8456147,8029555,Marra2019,10.1007/978-3-319-72389-1_43,  10.1007/978-3-642-16161-2_19,Zhang2008TowardAU,10.1145/2851613.2851797, Martinelli2019ObligationMI,10.1016/j.jnca.2011.03.019,Xu2007} and the frameworks proposed by \citet{5319024} and \citet{6045968} consider attribute \emph{updates} as a mechanism to trigger the re-evaluation of policies. 
\citet{Zhang2008TowardAU} support dynamic resource attributes (e.g., virtual machine storage and network bandwidth) that can affect the usage of data in the context of collaborative computing systems. 

\subsubsection{Detective.}
In addition to providing preventive mechanisms, the frameworks presented in \cite{Jung2014,Schtte2018LUCONDF,6045968,6405363,Wchner2013CompliancePreservingCS,Silva2010,6258293,9314823} also include detective mechanisms. The detective mechanisms proposed by \citet{Jung2014}, \citet{6045968}, \citet{6405363}, \citet{Wchner2013CompliancePreservingCS}, \citet{Silva2010}, and \citet{6258293} allow for the enforcement of actions performed by the PXP, the PEP, or the \emph{event signaler}. According to \citet{pretschner2006distributed}, execution actions are actions that typically involve internal interactions between system components.
\citet{6258293} group execution actions into four main types: logging; notifying a user; starting an activity; and faking information (e.g., a location).
In turn, \citet{9314823} enforce actions such as anonymizing data and deleting traces of data. 
While, \citet{Schtte2018LUCONDF} define detective enforcement as a statistic type of enforcement that acts as a formal audit support system, which provides evidence in the event of a policy violation or is used to manage policy conflicts. 

\subsubsection{Continuity of Enforcement.}
Continuity of enforcement involves re-evaluating a policy in order to test if the subject still complies with the policy rules and checking the validity of the continuous usage of an object, for instance checking for updates to attributes, context, or conditions and/or the fulfillment of obligations.
Most of the usage control frameworks address one or more of the following factors:
\begin{inlineroman}
\item monitoring attribute updates; \item evaluating the fulfillment of obligations  \item evaluating changes to environmental conditions, system conditions and/or contextual information. \end{inlineroman}
Depending on the specific framework, continuity of enforcement triggers the re-evaluation of a policy by: a PDP, a \emph{Prolog engine}, an \emph{evaluation engine}, a \emph{reference monitor}, a \emph{control monitor}, or a \emph{data usage transparency} component.
In cases where the framework supports the mutability of attributes, some components are used to observe and store attribute updates during an ongoing usage process. For instance, the frameworks presented in \cite{Lazouski2012APF,6470943,CARNIANI201637,10.1007/978-3-319-68063-7_8, 9161997,8717790,8456147,8029555,Marra2019,10.1007/978-3-319-72389-1_43,  10.1007/978-3-642-16161-2_19,10.1145/2851613.2851797,Martinelli2019ObligationMI} employ a \emph{usage monitor} or a \emph{session manager}, the framework proposed by \citet{Zhang2008TowardAU} uses an \emph{attribute repository}, whereas the framework proposed by \citet{10.1007/978-3-642-16161-2_19} utilises an \emph{evaluation engine}.
The remaining frameworks use a dedicated continuity of enforcement component in order to monitor updates to policy conditions and contextual information, as well as the fulfillment of obligations.
While, the framework proposed by \citet{Schtte2018LUCONDF} continuously monitors data flow by observing and controlling security breaches using a \emph{Prolog engine} and \emph{data flow tracking tools} (DFT).
The framework proposed by \citet{9314823} performs \emph{policy matching} of data consumer requests against data provider policies before allowing users to access the data. Although this framework does not apply continuous monitoring of data usage after access is granted, it does provide preventive execution actions performed via \emph{infrastructure services}, such as anonymizing data or orchestrating services that prevent users from performing certain actions on the data.

\subsubsection{Conflict Detection \& Resolution.}
A handful of papers \cite{CAO2020998,Schtte2018LUCONDF,Baldini2013,Neisse2015SecKitAM} specifically mention conflict detection and resolution. The \emph{data usage transparency} component proposed by \citet{CAO2020998} employs a defeasible reasoning engine that can detect and resolve conflicts that arise from usage control policies. However, the authors do not elaborate on the specific conflict resolution strategies that they employ.
In the case of the LUCON framework \cite{Schtte2018LUCONDF}, policies are compiled into Prolog programs. The authors highlight the importance of the employed logic based formalism as it allows for reasoning over policies in order to detect conflicting or incomplete rules. However, here again, the authors do not specify any conflict resolution strategies. 
In turn, the frameworks proposed by \citet{Baldini2013} and \citet{Neisse2015SecKitAM} uses XACML combining algorithms \cite{Florian2010}, such as \emph{permit-overrides}, \emph{deny-overrides}, or \emph{first-applicable} algorithms, in order to reach a decision when multiple rules return conflicting responses. 

\subsubsection{Administration.}
Interfaces that allow users to create and manage their usage control policies, modelled on XACML \emph{policy administration points}, are proposed by \cite {Lazouski2012APF,6470943,CARNIANI201637,10.1007/978-3-319-68063-7_8,9161997,8717790,8456147,8029555,Marra2019,10.1007/978-3-319-72389-1_43,Jung2014,10.1007/978-3-642-16161-2_19,10.1145/2851613.2851797,Martinelli2019ObligationMI,su12093885,MunozArcentales2019AnAF}.
Other frameworks propose \emph{graphical user interfaces} (GUI) \cite{CAO2020998,Baldini2013,Neisse2015SecKitAM,9314823}, simple policy editors \cite{Silva2010}, rely on operating system command line interfaces \cite{6258293,Xu2007}, or leverage editors that are built into development environments \cite{Schtte2018LUCONDF}.

\subsection{Robustness}
\begin{table} [t!]
\centering
\tiny
\caption{Robustness of Usage Control Mechanisms}
\begin{tabular}{p{2.2cm} p{2cm} p{1.8cm} p{2cm} p{1.5cm} p{2cm}}

\rowcolor[HTML]{C0C0C0} 
\cellcolor[HTML]{C0C0C0} &
  \cellcolor[HTML]{C0C0C0} &
  \cellcolor[HTML]{C0C0C0} &
  \cellcolor[HTML]{C0C0C0} &
  \multicolumn{2}{c}{\cellcolor[HTML]{C0C0C0}\textbf{Reliability}}
  \cellcolor[HTML]{C0C0C0}\\ 
\rowcolor[HTML]{C0C0C0} 
\multirow{-2}{*}{\cellcolor[HTML]{C0C0C0}\textbf{Framework}} &
  \multirow{-2}{*}{\cellcolor[HTML]{C0C0C0}\textbf{\begin{tabular}[c]{@{}l@{}}Performance \\ \& Scalability\end{tabular}}} &
  \multirow{-2}{*}{\cellcolor[HTML]{C0C0C0}\textbf{\begin{tabular}[c]{@{}l@{}}Interoperability \\ \& Compatibility\end{tabular}}} &
  \multirow{-2}{*}{\cellcolor[HTML]{C0C0C0}\textbf{Usability}} 
  &\textbf{Transparency} &
  \textbf{System Reliability} \\ 

\citet{10.1007/978-3-642-16161-2_19} &
  \begin{tabular}[c]{@{}l@{}} performance metrics\\ comparison \end{tabular}
  &XML 
  &\_
  &\_ 
  &\_\\ 

\rowcolor[HTML]{EFEFEF} 
\citet{Baldini2013,Neisse2015SecKitAM} &performance metrics &XML &use of templates
&trust &support for trust management\\ 

\citet{CAO2020998} &
  performance metrics &
  \begin{tabular}[c]{@{}l@{}}
  XML \\
  REST APIs 
  \end{tabular}
  &technical expertise required
  &\begin{tabular}[c]{@{}l@{}}
  explanations\\ provenance \end{tabular}
  &\_\\ 

   \rowcolor[HTML]{EFEFEF} 
\citet{6470943,CARNIANI201637} &performance metrics &XML 
&\_ &\_ &\_ \\ 

\citet{8456147} &performance metrics &XML &\_
 &\_ &\_ \\ 

   \rowcolor[HTML]{EFEFEF} 
\citet{9314823} &
  performance metrics &
  XML 
  &\_
  &\begin{tabular}[c]{@{}l@{}}
  audit\\ provenance \end{tabular}
  &decentralized policy enforcement \\ 

\citet{6258293} &\begin{tabular}[c]{@{}l@{}} performance metrics\\ other evaluation metrics \end{tabular} &XML
&\_
&trust &\_ \\ 

   \rowcolor[HTML]{EFEFEF} 
\citet{10.1007/978-3-319-68063-7_8} &performance metrics &XML & \_
&\_ &\_ \\ 
  
\citet{Jung2014}  
  &\_ 
  &XML 
  &technical expertise required
  &audit 
  &\_ \\ 

   \rowcolor[HTML]{EFEFEF} 
\citet{kateb2014} &\_  &XML  
  &\_
  &\_ &\_ \\ 

\citet{Lazouski2012APF} &
  performance metrics &
  XML 
  &\_
  &\_ & 
  \_ \\ 

   \rowcolor[HTML]{EFEFEF} 
  \citet{9161997,8717790} &performance metrics &XML 
  &\_
  &\_ &\_ \\ 
 
  \citet{8029555} &performance metrics &XML 
  &\_
  &\_ &\_ \\ 
  
  \rowcolor[HTML]{EFEFEF} 
  \citet{Marra2019} &performance metrics &XML 
  &\_
  &\_ &\_\\ 
  
  \citet{10.1007/978-3-319-72389-1_43} &performance metrics &XML
  &\_
  &\_ &\_ \\ 

   \rowcolor[HTML]{EFEFEF} 
\citet{10.1145/2851613.2851797} &\_
   & XML 
  &\_
  &\_
  &\_
   \\ 

\citet{Martinelli2019ObligationMI} &
  \_ &XML 
  &\_
  &\_ 
  &\_ \\ 
  \rowcolor[HTML]{EFEFEF} 
  
     \rowcolor[HTML]{EFEFEF} 
\citet{su12093885,MunozArcentales2019AnAF} &
  performance metrics &
  
  XML 
  &\_
  &trust & trusted technology stack
   \\ 

\citet{6045968} & performance metrics
  &
  XML 
  &\_
  & trust &\_ \\ 

   \rowcolor[HTML]{EFEFEF} 
\citet{5319024} &\_ &\_ 
  &\_
  &audit &\_ \\ 
  
\citet{Schtte2018LUCONDF} &
  performance metrics &
  Java 
  &technical expertise required
  & \begin{tabular}[c]{@{}l@{}}
  explanations\\ audit \end{tabular} &
   \_ \\ 

   \rowcolor[HTML]{EFEFEF} 
\citet{10.1016/j.jnca.2011.03.019} &
  performance metrics &
  \_
  &\_
  &\_ &
  \_ \\ 
  
\citet{6405363} &\begin{tabular}[c]{@{}l@{}} performance metrics\\ other evaluation metrics \end{tabular} &XML 
&\_
&trust  & reduced system complexity
\\ 

   \rowcolor[HTML]{EFEFEF} 
\citet{Wchner2013CompliancePreservingCS} &\begin{tabular}[c]{@{}l@{}} performance metrics\\ other evaluation metrics \end{tabular} &XML
&\_
&trust &\_ \\  
   
\citet{Silva2010} & performance metrics & XML
   &\_
   &\begin{tabular}[c]{@{}l@{}}
  audit\\ trust \end{tabular}
   &\_ \\ 
   \rowcolor[HTML]{EFEFEF} 

   \rowcolor[HTML]{EFEFEF} 
\citet{Xu2007} &
  \_ &
  XML 
  &\_
  &\_ &
  \_ \\ 
  
\citet{Zhang2008TowardAU} 
  &performance metrics 
  &XML 
  &\_
  &\_ 
  &\_ \\ 
 
\end{tabular}
\label{tab:robustness}
\vspace{-3mm}
\end{table}

Usage control robustness is an all-encompassing term used to refer to performance, scalability, interoperability, compatability, usability, transparency, and reliability. In the following, we elaborate on the various robustness mechanisms employed by the usage control frameworks that are summarized in Table \ref{tab:robustness}.

\subsubsection{Performance \& Scalability.}
The framework adopted by \citet{6470943} and \citet{CARNIANI201637} 
was originally employed in a cloud computing setting. 
In order to validate the original proposal, different extended versions of the same framework are evaluated in various use cases, such as, enforcing parental control using Smart TVS \cite{10.1007/978-3-319-68063-7_8};   
enhancing the security of Fifth Generation (5G) network systems \cite{9161997,8717790};
enforcing data protecting in industrial IoT settings, namely,
collaborative smart services \cite{8456147}, smart homes \cite{8029555}, general-purpose IoT architectures \cite{Marra2019}, and securing communications between IoT devices \cite{10.1007/978-3-319-72389-1_43}.
Moreover, another extension is mentioned in \cite{Martinelli2019ObligationMI} without any details on the physical implementation or performance.
Finally, the framework was used in a mobile computing context \cite{10.1145/2851613.2851797}, however, the authors do not provide any evaluation details.
The frameworks presented in \cite{CAO2020998,Schtte2018LUCONDF, 10.1007/978-3-642-16161-2_19,Zhang2008TowardAU, Lazouski2012APF, su12093885,MunozArcentales2019AnAF,9314823, 10.1016/j.jnca.2011.03.019,Baldini2013,Neisse2015SecKitAM,6045968,Silva2010,6258293,6405363,Wchner2013CompliancePreservingCS} are evaluated using various performance metrics, such as \emph{the time needed to evaluate a policy decision request}; \emph{memory usage}; or \emph{the time needed to enforce a policy decision}. 
While, the frameworks presented in \cite{CAO2020998,su12093885,MunozArcentales2019AnAF,9314823,Baldini2013,Neisse2015SecKitAM} are validated using different uses cases in the context of industry 4.0, particularly, for easing and ensuring secure data sharing in ecosystems, such as smart cities. 
Only the framework proposed by \citet{10.1007/978-3-642-16161-2_19} is evaluated in comparison to other usage control proposals. The authors compared the execution time of ConUCON to an existing security mechanism based on common actions that a user would perform on their phone.
In addition to using performance metrics to evaluate their approaches, the frameworks proposed by \citet{6258293}, \citet{6405363}, and \citet{Wchner2013CompliancePreservingCS} employ others measures that are relevant from a security perspective (e.g., analyzing attacker models and evaluating security countermeasures).
 
\subsubsection{Interoperability \& Compatibility.}
Most frameworks encode their policies using XML due to its strong interoperability capability.
Additionally, \citet{CAO2020998} have designed their framework components as an Application Programming Interface (API) in order to support the interoperability of shared services between data consumers and data providers that do not belong to the same domain. Although, the authors highlight the fact that semantic technologies could be used to facilitate exchange between users, they do not leverage semantic technologies in their framework.
In turn, \citet{Schtte2018LUCONDF} adopt the Java programming language, which supports different protocol adapters that are particularly suitable for IoT scenarios where data from different sources must be unified. Hence, the authors implemented  their PEP
using Apache Camel, which supports more than 240 protocols (e.g., Hypertext Transfer Protocol (HTTP), MQ Telemetry Transport (MQTT), etc.).

\subsubsection{Usability.}

The frameworks presented in
\cite{10.1007/978-3-642-16161-2_19,10.1145/2851613.2851797,Lazouski2012APF,10.1007/978-3-319-68063-7_8,8456147,8029555,Marra2019,Martinelli2019ObligationMI,su12093885,MunozArcentales2019AnAF,9314823}
do not discuss the usability of either their policy languages or their user interfaces. 
Whereas, \citet{6258293} mention that the usability of a policy language is always a concern. On the one hand, the expressiveness of a language can reflect the real use cases of usage control policies. On the other hand, the correct and adequate use of complex policies is hard for non-expert users.
The authors of the framework proposed by \citet{10.1016/j.jnca.2011.03.019}  claim that their policy language is an easy-to-use and well-defined language, but they do not provide any evaluation of the usability of their prototype. 
\citet{CAO2020998} provide an interface whereby users need to have basic knowledge with respect to defeasible logic in order to manage policies.
The GUI proposed by \cite{Baldini2013,Neisse2015SecKitAM} simplifies personal data management by providing a variety of policy templates that can be employed by users. Additionally, although the authors of the LUCON framework \cite{Schtte2018LUCONDF} claim that their language is easy to understand, users need to be  familiar with the UCON DSL grammar in order to write policies.
The interface proposed by \citet{Jung2014} can only be used by experts that know the various events and system actions that can be used within the policy description.
However, the authors mentioned that they are researching approaches on how to build user-friendly specification interfaces that allow even unskilled users to specify their security demands. They also indicated that the use of different usability patterns that include different user groups with varying skill levels and expertise  will be further studied.

\subsubsection{Reliability.}
The reliability of the usage control framework is highly dependent on different mechanisms employed in order to ensure the transparency and reliability of the system.
\paragraph{Transparency.} 
As depicted in Figure \ref{fig:taxonomy-requirements}, transparency can be established through four dimensions.
The first dimension for ensuring transparency is the use of auditing tools. Several authors \cite{Jung2014,Schtte2018LUCONDF,9314823, 6045968,6405363,Wchner2013CompliancePreservingCS,Silva2010,6258293,5319024} have emphasized the importance of using auditing mechanisms by including logging tools in their frameworks in order to provide evidence that users are using the data according to agreed policies, but also to track violations that usage control solutions are not able to detect (i.e., uncontrollable policies).
The second dimension relates to data provenance. \citet{CAO2020998} designed the \emph{data usage traceability} component based on defeasible logic for the purpose of tracing data usage history, whereas 
\citet{9314823} made use of blockchain technologies in order to ensure traceability of data usage.
The third dimension concerns explanations. Only two frameworks \cite{CAO2020998,Schtte2018LUCONDF} provide proofs to justify the decisions of the system
using inference engines based on defeasible and first order logics, respectively. Providing explanations of how the decision of either granting, denying, or revoking access was reached can help users to trust the decisions of the framework, but also support policy authors in fixing issues. 
The last dimension concerns trust. According to \citet{CAO2020998}, one important aspect of building trust is for the data owner to be able to exercise control over the usage of the data by other actors. \citet{su12093885,MunozArcentales2019AnAF} 
highlight the importance of using reliable and transparent usage control mechanisms in order for trust to be fully ensured. The frameworks presented in \cite{6045968,6405363,Wchner2013CompliancePreservingCS,Silva2010,6258293} employ \emph{trusted computing technologies}, which enhance usage control mechanisms through the inclusion of hardware-based trust components. 
Whereas, the reference architecture presented in \cite{su12093885,MunozArcentales2019AnAF}
relies on \emph{international data space} (IDS) \cite{Otto2018} connectors, which ensure that all the connectors or mechanisms involved in a data exchange run on top of a certified software stack. This is done through 
the IDS certification body that provides certifications for the connectors in order to establish trust among all participants. 
The framework presented in 
\cite{Baldini2013,Neisse2015SecKitAM} includes trust management by modelling trust relationships and recommendations using the Seckit policy language. 

\paragraph{System Reliability.}
The level of reliability  depends on the ability of usage control solutions to guarantee that usage control policies are enforced correctly.   
\citet{Baldini2013} and \citet{Neisse2015SecKitAM} relate the reliability of their usage control framework to the support for trust management that can ensure reliable and trusted relationships between architectural components and users. \citet{9314823} mention that  their choice of decentralized enforcement of policies improves the reliability of their framework by ensuring a good orchestration and synchronization between the system components. The framework proposed by \citet{su12093885,MunozArcentales2019AnAF} relies on IDS trusted connectors, which guarantee a reliable environment that enables usage control. \citet{6405363} claim that the reduction in complexity of their system compared to other usage control systems increases the reliability of their usage control framework. 
In turn, \citet{5319024} highlight the importance of adopting robust and reliable usage control frameworks. Although the authors do not consider or assess the reliability of their proposed framework, they do stress the importance of employing mechanisms that can protect usage control policies from unauthorized access, especially in distributed environments.
\section{Discussion}

In this section, we use the insights gained from our detailed usage control framework analysis in order to highlight open challenges and opportunities for the usage control domain in general and decentralized systems in particular.

\subsection{Gaps Analysis}
In the following, we identify open challenges and opportunities for the usage control domain, which we categorize under the headings: generality of policies; automated formal analysis, usability, verification and validation; and benchmarking. 

\subsubsection{Generality of Policies.} 
The majority of frameworks rely on specific policy languages that were developed according to domain-specific requirements in relation to \emph{Mobile, Cloud, IoT, and Industry 4.0.} (e.g., DUPO \cite{CAO2020998} and U-XACML \cite{10.1007/978-1-4419-6794-7_11}) and \emph {networking, operating systems, and collaborative software} (e.g., LUCON \cite{Schtte2018LUCONDF} and UCON\textsubscript{KI} \cite{Xu2007}). While, the policy languages that are meant to be \emph{domain-agnostic} are either not validated using use cases (e.g, OB-XACML \cite{kateb2014}) or are only evaluated in a specific domain (e.g., IND\textsuperscript{2}UCE \cite{Jung2014} and \cite{Silva2010}).
Hence, it is unclear if the existing proposals could be used for usage control in the general sense, where a single system may need to support privacy preferences, regulatory requirements, licenses, etc.
Given that usage control policies require both domain and application specific information, semantic technologies could potentially be used to develop a common policy model that provides support for different types of usage control policies. Semantic technologies are particularly suitable for the specification of policies, as ontologies and vocabularies can be used to formalize both policy concepts and rules in an extensible manner. 
In their privacy and data protection survey, \citet{estevesanalysis} examine various policy languages that leverage semantic technologies, such as ODRL \cite{ODRL2018}, the \emph{data privacy vocabularies} (DPV) \cite{Bonatti2018DataPV}, GDPRtExt \cite{Pandit2018GDPRtEXTG}, and the SPECIAL \emph{policy language} (SPL).
In another survey, \citet{Pellegrini2018} discuss how semantic technology based policy languages can be used to express DRM restrictions. Among the most prominent vocabularies are MPEG-21\footnote{MPEG-21, https://mpeg.chiariglione.org/standards/mpeg-21}, ODRL, the \emph{creative commons rights expression language} (ccREL)\footnote{ccREL, https://www.w3.org/Submission/ccREL/}, and XACML, to name but a few.
However, here also, it is unclear if the existing proposals are suitable for a system that needs to consider a variety of different usage control policies.

\subsubsection{Automated Formal Analysis.}  
The majority of existing usage control frameworks do not include a reasoning engine that could be used to automatically enforce usage control policies. This is usually due to the fact that their policy languages lack underlying formal semantics. 
According to \citet{HAN2012477}, formal approaches for describing policy languages facilitate automated analysis, policy verification, and explanations that can help manage the behavior of a system. For instance, automatic formal analysis could help to continuously verify compliance using usage policies, audit trails, and shared data. While, policy verification could be used to manage conflicting policies and to ensure policy consistency.
Growing dynamic environments, such as the web or IoT-based data sharing systems, where new users continuously join, pose new challenges in terms of unpredictability and dynamicity \cite{Dautov2018}. When it comes to automated policy analysis, the W3C community group\footnote{https://w3c.github.io/odrl/formal-semantics/} is working on a formal semantics for the ODRL standard, which could potentially be used to facilitate the automated analysis needed to cater for system unpredictability and dynamicity. Unfortunately, the provision of a usage control framework, which provides a blueprint for the development of an architecture that leverages said formal semantics, is outside of the remit of the ODRL community group.

\subsubsection{Usability.} 
Based on our analysis of various usage control frameworks according to application domain, we observe that the operating systems, collaborative computer systems, and network security domains do not focus on the usability of the interfaces that they provide. That being said, generally speaking usability is a consideration when it comes to mobile computing and Industry 4.0 as these areas attempt to empower users to manage their own data and how it is used by data consumers. 
Since the entry into force of the GDPR in 2018, significant changes have been observed in the way personal data are processed and the rights afforded to data subjects. 
The empowerment of users implies facilitating user awareness via tools that enable users to: \begin{inlineroman}
\item give their consent for personal data processing; 
\item provide preferences concerning how their data should be handled; \item benefit from data processing transparency \item profit from explainable policy decisions\end{inlineroman}. For this, administration interfaces must be user-centered and user-friendly, thus enabling the user to understand their rights while at the same time guiding them in the protection of their data. In the case of the frameworks examined herein, most of them do not mention usability and/or human-computer interaction (HCI) as important design considerations.
There is a broad body of literature that could potentially be used to enhance the usability of existing usage control systems. For instance, \citet{Mazumder2014USABILITYGF} propose usability principles that enable designers to consider the usability of a system early in the development cycle.
Whereas,
different works from privacy/legal researchers, such as \cite{kirraneSpecial2019,Drozd2020PrivacyCC}, could potentially be used to guide both usage control solution design and evaluation. 
\subsubsection{Verification and Validation.} 
Usage control policy-based frameworks are intended to protect data from malicious use and prevent unwanted operations such as the sharing data with untrusted third parties. Usage control policies are usually based on high-level goals provided by data providers/owners that are then translated into machine-readable policies. The success of a usage control system relies heavily on the absence of discrepancies between policy specification and their intended high-level goals. 
A major drawback observed after comparing the various usage control frameworks is the lack of verification and validation tools that can be used to assess the accuracy and the correctness of the proposed usage control mechanisms. 
When it comes to the broader verification and validation literature, there are a number of possible techniques, such as those examined by \citet{karafili2017verification}, which have already been successfully applied to systems that employ \emph{model checking}, \emph{algebraic solutions}, \emph{abductive reasoning}, \emph{answer set programming solvers}.

\subsubsection{Benchmarking.}  
As for evaluating the effectiveness of the various usage control solutions, the authors propose different evaluation schemes, which makes it difficult to compare the various proposals against one another.
For example, in \cite{Zhang2008TowardAU} the execution time metric is calculated  based on the time required to update policy attributes, to interpret the policy rules, and to communicate with other components; whereas in \cite{10.1007/978-3-642-16161-2_19} the metric 
depends on the type of actions performed by the users and the time needed to perform obligations and retrieve contextual or attribute information.  Consequently, there is a need to develop a usage control benchmark that defines standardized characteristics, such as performance, scalability, etc.
In their work, \citet{Kistowski2015} give important insights on how to build standardized benchmarks  based on a set of quality criteria. The authors cite \emph{relevance}, \emph{reproducibility}, \emph{fairness}, \emph{verifiability}, and \emph{usability}  as key characteristics needed to build a benchmark. In our case, \emph{relevance} implies choosing suitable and applicable characteristics for comparing different usage control frameworks. Based on our comparative analysis, we have already observed some common metrics for evaluating usage control frameworks, for instance, the time to process usage control policies and the memory usage needed to process policies.
The remaining features imply that the benchmark is capable of: \begin{inlineroman} \item reproducing similar results when the same test configuration is used; \item ensuring fair competition with other benchmarking tools; \item producing reliable and accurate results \item providing user-friendly tools that facilitate the execution of comparison experiments \end{inlineroman}. 

\subsection{Decentralized Usage Control}
Decentralized IoT-based use case scenarios, such as the motivating scenario which we introduced in Section 2, bring an additional set of considerations from a usage control perspective. Thus, we end our analysis with a discussion of the various usage control challenges and opportunities that have been derived from our motivating use case scenario. 

\subsubsection{Data sharing and Lack of Control.}
Once data are shared or accessed, they will move outside the premises of the data provider and therefore out of their control. For instance, users of the data sharing platform in our IoT-based scenario, will no longer be able to control what happens to their data after they give their consent for stakeholders to use this data.
Besides, different copies and derivations of the same data can be shared across the network, which makes it difficult to control how data are used \cite{6405363,6258293}.
When it comes to the literature with respect to increasing control over data usage, proposals either involve employing \emph{information flow tracking tools} or using \emph{sticky policies} \cite{10.1145/3447867,Demir2016SurveyOA,Miorandi2020StickyPA}.

\paragraph{Information Flow Tracking Tools.}
Dynamic \emph{information flow tracking} (IFT) involves the tagging and tracking of data as they propagate across network systems \cite{10.1145/3447867}. In distributed usage control, IFT is deemed complementary to policy enforcement, as it allows for the protection of the different data derivations shared across system nodes in a distributed environment \cite{iFlorian2015,6405363,6258293,10.1007/978-3-642-28879-1_9,10.1145/3183342}. 
IFT tools are often tailored for scenarios such as the web \cite{Janic2013TransparencyET,10.1007/s12394-010-0069-4,Yu2017EffectivelyPY} and the  cloud \cite{kunz2020towards,7987219,Pappas2013CloudFenceDF}, among others, in order to cater for different aspects of data usage control, namely privacy preservation, regulatory compliance, and sensitive information protection. 
Generally speaking, IFT tools are either hardware-based or software-based. According to \citet{10.1145/3447867}, most of the hardware IFT solutions can be vulnerable to security threats and are complex to debug. While, \citet{Demir2016SurveyOA} point out that most IFT software solutions come with low security guarantees and high programmability impacts, which increases the reluctance to use such techniques.

\paragraph{Sticky Policies.}
Sticky policies represent restrictions on the use of data, which are directly attached to the corresponding data. Interest in sticky policies has increased with the emergence of new distributed technologies \cite{Miorandi2020StickyPA}. By using sticky policies in a decentralized environment, data consumers could potentially ensure that their usage policies are enforced after data are transferred from one system to another. In their survey, \citet{Miorandi2020StickyPA} compare and contrast sticky policy solutions for the cloud, the IoT, and context-aware applications, among others. Additionally, the authors outline several approaches for sticky policies, such as, encryption (e.g., public-Key encryption (PKE), identity-based encryption (IBE), and attribute-based encryption (ABE)), sticky policy languages (e.g., the enterprise privacy authorization language (EPAL) and the PrimeLife privacy policy language (PPL)), and sticky policy for access control (e.g., the purpose-aware role based access control (PuRBAC) model). They also provide a list of challenges that still need to be addressed before sticky policies can be adopted. For instance, the need for robust encryption techniques; the standardization of an expressive sticky policy language; and the adoption of mechanisms that are able to distribute and synchronize policies across decentralized systems in a controllable and secure way.

\subsubsection{Distributed Trust Management.}
A distributed environment can be composed of heterogeneous entities and systems that interact with each other, which brings about trust issues in relation to \emph{dependability}, \emph{security}, and \emph{reliability} \cite{ahmed2019TrustAR}.
Accordingly, in our IoT scenario,  
smart objects  (e.g., sensors, phones, cars, and services) and entities (e.g, the marketing company, users, and manufacturers) can be of different forms and types, hence, the interaction between the different entities and devices  should be trusted in order to ensure secure and reliable infrastructure.
In their survey, \citet{Artz2007ASO} compare and contrast different approaches for trust management in computer science, namely, policy-based trust, reputation-based trust, and general models of trust that could serve as interesting starting points.
The establishment of trust can be facilitated via the use of trust negotiation solutions that are based on enforcing trust policies and/or the exchange of security credentials, among others. Different semantic languages have already built-in mechanisms for trust negotiation. For instance, the \emph{web services security} (WS-Security) policy\footnote{WS-Policy, https://www.w3.org/Submission/WS-Policy/}, a standard introduced by the W3C, in order to provide ways to attach signatures and encryption headers or security tokens during communication between system nodes, or the PeerTrust language \cite{10.1007/978-3-540-30073-1_9} that involves automated trust negotiation.
The exchange of credentials implies the presence of an \emph{authority} that can verify the provided credentials. An example is the \emph{international data space} (IDS) \cite{Otto2018} initiative that specifies that all parties involved in data exchange need to use IDS certifications issued by the IDS certification body, which is a trusted third party.
Another example concerns the use of blockchain smart contracts, where there is no need for a third party to verify identities. Smart contracts are very relevant in fields such as the IoT, where different parties are involved in data exchange, as they can be used to facilitate trusted and reliable transactions \cite{8048663}. Several proposals that employ smart contracts in order to establish trust have already been proposed (cf., \cite{WU2021101330,ESPOSITO2020102308,9120287,9142395}). 
When it comes to other approaches to trust management, reputation-based trust can be used to assess trust relationships, while general models of trust involve trust properties and relationships from different domains with different resources \cite{Miorandi2020StickyPA}. That being said, 
distributed applications impose challenges, such as the need for scalable trust management solutions that support an increasing number of nodes; the importance of considering different types of reputation metrics  that vary from one domain to another; and the automation of trust management using trust policies.

\subsubsection{Distributed Enforcement.}  
In a usage control scenario, two main entities are always present: \emph{the data provider} and \emph{the data consumer}. In our IoT use case, the data consumer and data provider roles may be interchangeable, as a data consumer may be a data provider, and vice versa. For example, manufacturers can be data consumers of residents' data. However, they can also be providers of residents' data to other stakeholders, such as the marketing companies.
Hence, a distributed system must not only enforce usage control policies on data leaving the providers domain, but also control the usage to resources or data in the consumers domain. 

\paragraph{Policy Enforcement Point and Policy Decision Point.}
Using  IFT to track data  or sticky policies to enforce usage policies, or a combination of both, requires techniques that are capable of ensuring that policies and data usage are always observed and enforced in a decentralized environment. \citet{iFlorian2015} and \citet{9314823} highlight the importance of using a reference architecture including a local policy enforcement point and a policy decision point at every site. 
In addition, mechanisms such as auditing should be put in place in order to trace the various interactions and transactions that occur within a system, but also to collect information provided by data monitoring tools.

\paragraph{Auditing.}
\citet{pretschner2006distributed} state that detective mechanisms, particularly audit logs, are of particular interest in a usage control framework in order to verify that data consumers abide by usage control policies. Detective mechanisms are complementary to preventive mechanisms, especially if  data are shared with third parties and the control of how data are used becomes more difficult. Moreover, according to \citet{Bonatti2017}, audit logs present the core of any transparency architecture as they can be used to both record data transactions and what happened to the data.
%
In a decentralized usage control framework, the use of a hybrid approach involving a local internal ledger and a distributed ledger could be warranted. For instance, a local ledger could provide information about what happened inside the local component of the framework, while a distributed ledger could be used to trace  what happened between services/ components and their relationships. However, this approach presents different challenges when it comes to the interoperability between ledgers from different consumers, and difficulties in retracing all events in a unified manner, which complicates compliance checking.
Modeling events using semantic technologies could potentially resolve these issues by:
\begin{inlineroman} \item using a common schema to describe logs \item modeling and describing what happened to data using description vocabularies that represent and relate information from different sources (i.e., data consumers) in an understandable and timely manner\end{inlineroman}. There is already numerous vocabularies that could potentially be adapted to describe the provenance of data, such as \emph{PROV}\footnote{PROV, https://www.w3.org/TR/prov-overview/}. Additionally, there are different vocabularies to describe the data processing events, such as \emph{Event}\footnote{Events, http://motools.sourceforge.net/event/event.html} and the \emph{LODE} ontology\footnote{LODE, http://linkedevents.org/ontology/}. 

\section{Conclusion}
This paper provides an overview of the usage control domain by comparing and contrasting the predominant usage control frameworks found in the literature.
We started by examining various usage control concepts and providing a broad definition for usage control policies based on deontic concepts. 
Guided by an integrative research methodology, 
we collected and categorized requirements that have been used to guide the development of various usage control solutions. These requirements address both the ``what'' and ``how'' aspects  of  enforcing usage control policies, in particular, the specification, the enforcement, and the robustness of the proposed solution.
We subsequently compared and contrasted the predominant usage control frameworks based on our resulting taxonomy of requirements.

We subsequently broadened the discussion to include opportunities and challenges that can guide future research directions. We highlighted that due to their flexibility and interoperability, semantic technologies are particularly suitable for encoding usage control policies. Following on from this, we discussed the key role played by reasoning when it comes to policy compliance, consistency, transparency, and system security. Additionally, we identified the need for additional research from a usability and a human computer interaction perspective in order to enable users to manage their data. We also highlighted the importance of benchmarking in order to systematically assess and validate the robustness of different usage control solutions. Finally, we showed the need to employ verification and testing tools in order to enhance the reliability and accuracy of usage control proposals. 

Finally, we outlined further challenges and opportunities that arise with the emergence of modern decentralized and distributed environments. Key considerations include the potential brought about by information flow tracking tools and sticky policies when it comes to tackling issues with respect to controlling how data are shared and used. Additionally, we highlighted the key role played by distributed trust management approaches in order to establish dependable, secure, and reliable data sharing infrastructures. Finally, we underlined the need for preventive and detective enforcement mechanisms for ensuring the correct enforcement of usage policies.


\bibliographystyle{ACM-Reference-Format}
\bibliography{library}
\end{document}